\documentclass{article}

\usepackage[utf8]{inputenc} 
\usepackage[T1]{fontenc}    
\usepackage{hyperref}       
\usepackage{url}            
\usepackage{booktabs}       
\usepackage{amsfonts}       
\usepackage{nicefrac}       
\usepackage{microtype}      
\usepackage{xcolor}         
\usepackage{graphicx}
\usepackage{multirow}
\usepackage{enumitem}
\usepackage{comment}
\usepackage{amsmath}

\usepackage{caption}
\usepackage{subcaption}

\def\Dcal{\mathcal{D}}

\DeclareMathOperator*{\argmax}{arg\,max}

\author{
  Guangmingmei Yang, Xi Li, Hang Wang, David J. Miller, George Kesidis\\
School of EECS\\
Pennsylvania State Univ.\\
  University Park, PA 16802 \\
\texttt{gzy5102,xzl45,djm25,gik2@psu.edu} \\
}

\title{CEPA: Consensus Embedded Perturbation for\\
Agnostic Detection and Inversion of Backdoors}

\begin{document}

\maketitle

\begin{abstract}
A variety of defenses have been proposed against Trojans planted in (backdoor attacks on) deep neural network (DNN) classifiers. Backdoor-agnostic methods seek to reliably detect and/or to mitigate backdoors irrespective of the incorporation mechanism used by the attacker, while inversion methods explicitly assume one.
In this paper, we describe a new detector that: relies on embedded feature representations to estimate (invert) the backdoor and to identify its target class; can operate without access to the training dataset; and is highly effective for various incorporation mechanisms (i.e., is backdoor agnostic). 
Our detection approach is evaluated -- and found to be favorable -- in comparison with an array of published defenses for a variety of different attacks on the CIFAR-10 and CIFAR-100 image-classification domains. 
\end{abstract}

\section{Introduction} \label{sec:intro}

Deep learning is based on very large labelled training datasets. Poisoning of the training set through
insecure supply chains can result in the planting of backdoors (Trojans) in a deep neural network (DNN) classifier. Data poisoning can also occur during a model 
refinement process through insecure Reinforcement Learning with Human Feedback (RLHF), i.e., via an insider
or by a man-in-the-middle between the vendor and customer.
Backdoor attacks involve the choice of the backdoor pattern and the method of incorporating the backdoor pattern into poisoned training samples and, operationally, into test samples (called ``backdoor triggers'').  In dirty-label attacks,
clean (natural, correctly labeled) samples are drawn from one or more classes (called ``source classes''), the backdoor pattern is incorporated into them, and 
these samples are then mislabeled to a ``target class'' of the attack
\cite{BadNet}. The target class is the same as the source class, i.e., no mislabeling, in the case of so-called ``clean-label" attacks \cite{Madry-clean-label}. 
The poisoned samples are then added to the training set used to learn the classifier. 
For a successfully backdoor-poisoned classifier, presence of the backdoor trigger in a test sample will, with high probability, cause an attacker-desired incorrect class decision to be made.
Note that these attacks, which are successful against classifiers, can also be applied to large language models (LLMs) -- e.g.,
the different sentiments (e.g., `positive', `negative',`neutral') of the responses of a generative LLM can be thought of as classes, and the censorship mechanism of an LLM is also a kind of binary classification (with the LLM's response either censored or uncensored).
Backdoor attacks can also target models used for, e.g., translation,
summarization, regression or prediction --
for these tasks, presence of the backdoor trigger in a test pattern will result in the model producing a highly erroneous output/prediction.  

It has been experimentally observed that
dirty-label
attacks are highly 
effective even when the fraction of the training set that is poisoned is very low;  moreover, this is achieved without relying on over-training \cite{Belkin19}.  A backdoor requires much less computation to operationally trigger than an adversarial input (a.k.a. a test-time evasion attack), e.g., \cite{CW} -- 
the latter does not rely on data poisoning and 
thus requires the solution of an optimization problem to find a perturbation of the input pattern that induces the model to make an error.
Also note that ``universal" adversarial perturbations, e.g., \cite{Dezfooli,Hendrycks-natural,suffix23}, are a kind of ``natural" or ``intrinsic" backdoor.

Backdoor attacks on classifiers can also be categorized
based on the target class and source class(es) involved.
E.g., an all-to-one attack 
is such that all non-target classes are source classes (i.e., with some poisoned training samples).
Likewise, an ``all-to-all" attack is one where all classes are both source {\it and} target classes, 
where for each source class there is an attacker-designated target class
(i.e., a (source, target) class pair), and where each class pair could involve a unique backdoor trigger pattern.  There are also one-to-one 
attacks involving a single (source, target) class pair.

Backdoor attacks can further be categorized based on the method of incorporating the backdoor pattern into a given sample.
Various published attacks include 
patch replacement (BadNet) \cite{BadNet}, blending \cite{NC}, additive \cite{TNNLS}, and warping (WaNet) \cite{WaNet}. 
For ``digital'' attacks, where the attacker 
may make arbitrary changes to a poisoned training sample or to a (triggered)
test sample, there are in fact {\it myriad} ways the backdoor pattern can be incorporated into a sample\footnote{A physical attack on, e.g., image classification involves inserting a physical object (the backdoor pattern object) into a scene that is then imaged.  For digital attacks, the attacker has access to already captured digital images, with full autonomy to alter images to incorporate a (hopefully inconspicuous) backdoor pattern.  Thus, for digital attacks there are uncountable ways 
that can be used to incorporate a backdoor pattern.}.
The backdoor pattern could alter only a small number of features (e.g., a small number of pixels, in the case of image classification), or it may be
``global'', altering (but perhaps with tiny modifications to) {\it all} features comprising a given sample.
Backdoor triggers that are inconspicuous are more effective, as they 
make more challenging the defense tasks of backdoor detection, cleansing of the poisoned samples from the training set, as well as 
test-time trigger detection. 
In the case of image classification, ``inconspicuous'' triggers include both imperceptible backdoor triggers as well as perceptible triggers that are innocuous in the given scene (e.g., a bird in the sky).
Clean-label attacks may require either muting source-class discriminative features (e.g., by adding noise) or high poisoning rates in order for the backdoor pattern to be well-learned by the classifier; 
so, even though these attacks do not involve mislabeling, they  
may be more conspicuous than dirty-label attacks.

Various types of backdoor defenses have been proposed. They can be categorized in different ways, e.g., depending on:
{\it when} the defense is enacted (before/during training, post-training, during test/operation time);
whether the defense simply detects backdoor poisoning or, beyond this, mitigates/corrects (poisoned training samples, the poisoned model, or the decisions made operationally on backdoor-trigger test samples); and 
whether the defense is intrinsically {\it agnostic to} or relies on 
knowledge of or an assumption about the backdoor pattern incorporation mechanism -- the latter are often referred to as ``inversion" or ``reverse-engineering" defenses. Post-training and test-time defenses typically do not rely on access to the training set, but may leverage a small, clean (correctly labeled, unpoisoned) dataset (which is 
sufficient for supporting accurate backdoor detection, but which is wholly too small 
to be used for retraining an accurate DNN from scratch).
The post-training scenario is important because in practice the defender may not have access to the data set that was used to train the classifier/model.  For example, the model may have been purchased by an entity (e.g., a company or government), without any ``data rights'' allowing access to the training set that was used.
Moreover, if the model was built long ago (a legacy system),
the training set may be lost or inaccessible.

E.g., \cite{SS,AC} 
attempt to cleanse the training set, while
Fine-Pruning \cite{KLiu18} and I-BAU \cite{I-BAU} are both post-training mitigation approaches that refine the presumed poisoned model.
These four methods are agnostic to the manner
of backdoor incorporation.
Neural Cleanse (NC) \cite{NC} is a post-training, reverse-engineering detector that assumes 
the backdoor pattern 
is incorporated via a ``blending'' operation (with patch-replacement a special case of blending). \cite{NC} also proposed a method to mitigate the poisoned model, informed by the reverse-engineered backdoor trigger.
Related work is further discussed in the Appendix Section \ref{sec:prior}.

In this paper, we focus on 
inversion based  defense, which can be used to: detect whether or not a model was 
backdoor-poisoned (and, if so, identify its target class), to cleanse the training set,
e.g., \cite{SS,AC},
and for test-time backdoor-trigger detection.
Moreover, inversion defenses are fundamental to explainable/interpretable AI -- 
they identify a suspicious ``cause'' (the trigger pattern) for why the model is making a particular (perhaps peculiar) decision on a given test sample.
A weakness in most existing 
inversion defenses is that they may fail when the backdoor incorporation mechanism assumed by the defense
(e.g., additive) does not match the mechanism actually used by an attacker (e.g., blending).
This work remedies this weakness, developing a
post-training detector 
that reverse-engineers the backdoor pattern while, at the same time, being {\it agnostic} to the method of backdoor pattern incorporation.
Our attack-defense experiments, focused on the image classification domain, demonstrate that this approach,
with minimal assumptions,
achieves strong performance (a distinctive detection signal) across a variety of backdoor attacks with different incorporation methods.

\section{CEPA Detection \& Inversion Method}
\subsection{Backdoor Detection}
Let $p_t(x)$ be the DNN's posterior probability for class $t$, given input sample $x$, obtained via a softmax in the output layer of the network.
The classifier uses a winner-take-all-rule, i.e.
the decided class is $\hat{c}(x) = \argmax_k p_k(x)$.
Furthermore, given the feedforward DNN structure, the
class posterior can be expressed as
$p_t(x)=g_t(f(x)),$
where $f(x)$ is the output of some internal layer of the DNN, i.e., it is an embedded feature vector
activated by input sample $x$.

For each putative backdoor target class $t$ 
or, to account for possible one-to-one attacks, each putative target class $t$ paired with a putative source class $s\not = t$,
we formulate an optimization problem, seeking sample-wise input perturbations ($\delta_x$) which: 1) induce misclassification to the putative target class $t$; 
2) are inconspicuous (small), in some well-defined sense;
and 3) 
induce a {\it common} perturbation of the embedded feature vector $f(\cdot)$ (which is a putative backdoor pattern).
A backdoor detection can then be made if: 1) the $\delta_x$ have {\it e.g.} an unusually small average norm across all samples, denoted as $\|\Bar{\delta}\|$,
2) the induced embedded feature vector perturbations have unusually small variation about their mean ($\mu$) (with this mean treated as the common backdoor pattern in the embedded space), and/or 3) the common perturbation vector $\mu$ has unusually {\it large} magnitude (consistent with the backdoor overfitting phenomenon observed in \cite{MMBD}).
Likewise, for class pairs ($s,t)$ {\it not} involved in a backdoor attack, the estimated
input perturbations $\delta_x$ should essentially amount to 
 {\it adversarial perturbations},
 e.g., \cite{CW}.
 For successful 
 adversarial perturbations
 on two distinct input samples $x$ and $x'$ from the same class, there is {\it no} expectation that they will induce a {\it common} perturbation $\mu$ in the embedded space, i.e., we would expect that the variance
 about $\mu$ in the embedded space will be high for non-backdoor class pairs.  Moreover,
 for such pairs, since there is no backdoor mapping, there will be no ``backdoor overfitting'', i.e., we would expect that for such pairs $\|\mu\|$ will be small compared to that for the backdoor class pair.
Finally, via backdoor data poisoning, the model may learn to recognize {\it very subtle} patterns 
(e.g., a single-pixel modification, or minute per-pixel changes).  
By contrast, much larger input changes may be required for a successful TTE attack (inducing a targeted class decision change).

Accordingly, specialized for one-to-one attacks,
for a given putative (source, target) class pair $(s,t)$,
we define the objective function to be minimized over
the sample-wise input perturbations $\delta_x$\footnote{While $\delta_x$ {\it is} a function of the class pair $(s,t)$, we omit making this dependence explicit for notational concision.} and the common embedded perturbation $\mu$:
\begin{align}
\frac{1}{|\Dcal_s|}\sum_{x \in \Dcal_s} &
\Big(   - \log p_t(x+\delta_x)      \label{objective}\\
~ & + \lambda_1 \| f(x+\delta_x)-f(x) - \mu \|^2
+ \lambda_2 \| \delta_x\| \Big),
\nonumber
\end{align}
where
$\Dcal_s$ is the set of clean samples originating from class $s$,
$\lambda_1 \geq 0, \lambda_2 \geq 0$, the second term is the variance about the common embedded perturbation ($\mu$)\footnote{For
all-to-one attacks, replace $\Dcal_s$ by $\cup_{s\not= t}\Dcal_s$ in \eqref{objective}, and the following
$\mu,\sigma^2$ quantities will be indexed by (i.e., depend on) just $t$, not $s$ and $t$.}, and the third term is the average input perturbation norm.
Moreover, in practice we have found that setting $\lambda_2=0$ is effective, and use this setting in all reported experiments.

This objective function is iteratively minimized, where each iteration
consists of three steps: 
\begin{itemize}
\item multiple gradient descent steps taken w.r.t. $\delta_x, \forall x \in \Dcal_s$,
\item closed-form update of $\mu$,
noting that, given the input perturbations fixed, the globally
optimal $\mu_{s,t}$ is:
\begin{align}\label{eq:mu-def}
\mu_{s,t} & = \frac{1}{|\Dcal_s|}\sum_{x\in\Dcal_s} (f(x+\delta_x)-f(x)) .
\end{align}
Moreover, after each gradient descent step, $\delta_x$ is ``clipped"
so that $x+\delta_x$ is feasible, e.g., each element
in $[0,1]$ for the case of normalized pixel intensities. 
Note that by setting $\lambda_2=0$, we do not explicitly penalize the $\|\delta_x\|$ magnitudes. 
However, we implicitly minimize/limit $\|\delta_x\|$ sizes by initializing each $\delta_x$ to a zero vector and by (promptly) terminating the algorithm once both i) a sufficient fraction (e.g., 90\%) of source-class samples are misclassified to the target class and ii) no further improvement in embedded consensus (smaller variance) can be achieved.
\item The dual variable $\lambda_1$ is dynamically
adjusted by increasing/decreasing
its value by a factor ($1 + \alpha$ with $|\alpha|<1$), or by grid search,
to achieve the maximum consensus (minimum variance) possible, while still maintaining high induced misclassification fraction to the target class $t$.
In this sense, $\lambda_1$ is not truly a hyperparameter of the algorithm (this is discussed further below). Also, while
for detection we set $\lambda_2=0$, for reverse engineering (discussed further in section \ref{sec:backdoor-inversion})
we set both $\lambda_1$ and $\lambda_2$ greater than zero. 
\end{itemize}

Once the optimization has been performed for all candidate class
pairs $(s,t)$,
detection inference can be based on one of several detection statistics that involve the consensus embedded perturbation. In particular, backdoor detection if there is 
a class pair $(s,t)$ with 1) unusually small common (average) input perturbation norm $\|\Bar{\delta_{s,t}}\|:=\frac{1}{|\Dcal_s|}\sum_{x \in \Dcal_s} \|\delta_x\|$ or unusually high embedded perturbation {\it consensus} compared with other class pairs, characterized by: 2) common embedded perturbation $\mu_{s,t}$ with unusually {\em large} norm and/or 3) unusually small normalized standard deviation: 
$$\frac{\sigma_{s,t}}{\|\mu_{s,t}\|} ~:=~ \frac{\sqrt{\frac{1}{|\Dcal_s|}\sum_{x \in \Dcal_s} \| f(x+\delta_x)-f(x) - \mu_{s,t} \|^2}}{\|\mu_{s,t}\|}.$$
Detections can then be made based either on MAD \cite{NC} or on order-statistic p-values \cite{TNNLS}, assessed using a null model that is estimated using all (source, target) ($(s,t)$) class pair detection statistics. For example (and as applied to report our results here), if one or more of the three metrics $\|\Bar{\delta_{s,t}}\|$, $\|\mu_{s,t}\|$, and $\frac{\sigma_{s,t}}{\|\mu_{s,t}\|}$ for a potential (source, target) pair exceed the MAD threshold, we detect that the current class pair ($(s,t)$) has been backdoored. Conversely, if no metric exceeds the MAD threshold, we determine that the class pair is clean. Additionally, note that we have postulated that backdoored classes demonstrate {\it smaller} $\|\Bar{\delta_s}\|$ or $\frac{\sigma_{s,t}}{\|\mu_{s,t}\|}$ values and/or
{\it larger} $\|\mu_{s,t}\|$ values. 
Thus, anomalies should be assessed not only considering deviation from a central value (mean or median) but also by the {\it direction} (sign) of the deviation. 
Thus, for $\|\mu_{s,t}\|$, a backdoor is only identified when a MAD score for some class pair exceeds the threshold and the metric value is larger than the median.  Likewise, for
$\|\Bar{\delta_s}\|$ and $\frac{\sigma_{s,t}}{\|\mu_{s,t}\|}$, detections are only made if a MAD score exceeds the threshold and the metric value is smaller than the median.
Considering both deviations and their direction enhances the robustness of our detection method.

\subsection{Backdoor Inversion}\label{sec:backdoor-inversion}
Suppose that a backdoor attack has been detected involving (source, target) class pair $(s,t)$. 
To reverse-engineer (estimate) backdoor patterns in the input space, one should seek input perturbations that: 1) are small in size (consistent with the attack's imperceptibility), 2) lead to misclassification from the source class $s$ to the target class, $t$, and 3) collectively yield consensus in the embedded space (to ensure they are consistent with the embedded space (estimated)  backdoor mechanism). 
Accordingly, CEPA backdoor pattern inversion is based on {\em jointly} achieving:
\begin{itemize}
\item high misclassification rate 
\item high embedded perturbation consensus {\em and}
\item low mean perturbation norm, $\|\Bar{\delta_{s,t}}\|$.
\end{itemize}
Therefore, the reverse engineering objective is the same as the detection objective in (\ref{objective}) except that now we set $\lambda_2>0$. This explicitly constrains the input perturbation norms, the necessity of which has been empirically validated.

Similar to the detection process, we minimize the objective function over the $\delta_x$ and $\mu_{s,t}$. The stopping condition for the reverse engineering optimization is twofold: i) achieving a high misclassification rate to the target class and ii) reaching a point where no smaller input perturbation can be achieved.

Our (sample-dependent) estimates of the backdoor in the input space are given by the optimized sample-wise perturbations ${\delta}_x, \forall x \in \Dcal_s$.
Again, note that $\lambda_1$ and $\lambda_2$ are restricted to values strictly greater than zero throughout the optimization. 
This is because for the true backdoor target class, we observe that the CEPA optimization quickly reveals the backdoor ``shortcut'', without any need for Lagrangian multiplier adjustment. Moreover, we have found that any fixed positive $\lambda_1$ and $\lambda_2$ that result in small optimized perturbations({\it e.g.}, average norm smaller than 1) yield good estimates of the backdoor pattern.
Alternatively, it is also possible to set $\lambda_1$ and $\lambda_2$ dynamically: if the current estimates are too large (e.g., average norm larger than 5), decrease $\lambda_1$ or increase $\lambda_2$ until the estimates no longer saturate, and then fix the $(\lambda_1,\lambda_2)$ values for the subsequent optimization steps. 

Finally, CEPA reverse engineering can be coupled with other (non-CEPA) detection methods (which determine the detected pair $(s,t)$), e.g., MMBD \cite{backdoor-universal}. 
See Table \ref{tab:detection_accuracy} and
Section \ref{sec:backdoor-inv-performance} below.

\subsection{Discussion: Hyperparameters}

One might consider the internal layer $f(\cdot)$
and the Lagrange multiplier $\lambda_1$ to be hyperparameters 
of our detector.  However, our detection approach can be simultaneously applied to 
{\it various} layers of the DNN, with detection inference then based only on the layer with the {\it most extreme} detection
statistic (smallest p-value).  Thus, the internal layer can be chosen in an automated fashion.  
Similarly, as discussed earlier (and also in the experiments
section), $\lambda_1$ can also be automatically adjusted,
and is thus also not truly a hyperparameter.

For backdoor inversion, where $\lambda_1$ and $\lambda_2$ are 
strictly
positive values, again, 
we have found that accurate inversion is not very sensitive to these choices.  Moreover, their values can be dynamically selected.
The threshold on the misclassification rate (used to terminate the CEPA optimization) is a hyperparameter of the algorithm.  However, we have found experimentally (see the Appendix) that CEPA's performance is largely insensitive to the choice of this threshold.



\section{Experimental Results}

\subsection{Experiment Setup}\label{sec:experiment-setup}

\textbf{Datasets}: 
Our experiments are conducted on the benchmark datasets CIFAR-10 and CIFAR-100 \cite{cifar10}.
CIFAR-10 contains 60,000 $32\times32$ color images from 10 classes, with 5,000 images per class for training and 1,000 per class for testing.
CIFAR-100 contains 60,000 $32\times32$ color images from 100 classes, with 500 images per class for training and 100 per class for testing.\\
\textbf{Attack Settings}:
On CIFAR-10, we consider the following all-to-one
attacks:
1) additive global chessboard
\cite{TNNLS};
2) additive 1-pixel pattern \cite{SS} \cite{AC};
3) BadNet random patch \cite{BadNets};
4) BadNet unicolor patch \cite{NC};
5) local blend with random patch trigger \cite{Targeted-Backdoor};
6) global blend with Hello Kitty trigger \cite{Targeted-Backdoor};
7) WaNet trigger \cite{WaNet}.
On CIFAR-100, we just considered the BadNet attack.
Details of the backdoor patterns and attack configurations can be found in Appendix \ref{sec:setup_details}.
\\
\textbf{Training Settings}:
We trained ResNet-18 \cite{ResNet} models on CIFAR-10 and VGG-16 \cite{VGG} models on CIFAR-100. Also, for the WaNet attack, the PreActResNet-18 architecture was used, consistent with the authors' advice.
All the backdoor attacks we generated were highly successful, demonstrating a high attack success rate (ASR) with minimal impact on clean test-sample accuracy (ACC). Training details are shown in Appendix \ref{sec:setup_details}.\\
\textbf{Detection Settings}:
For CIFAR-10, we randomly selected 10 images per class (90 images in total) from the test set to form the small clean dataset possessed by the defender.
The remaining test instances were reserved for performance evaluation.
For CIFAR-100, we randomly selected 2 images per class (198 images in total) to form the small clean dataset.
For each putative target class,
the parameter $\lambda_1$ of \eqref{objective} was initially 
set to $0.000001$ and dynamically adjusted based on
the misclassification rate $\pi$ and misclassification threshold $\pi_0=0.9$:  if $\pi \geq \pi_0$ for 5 consecutive iterations, $\lambda_1$ is increased by a factor of 1.2
(reducing the variance around the current $\mu$).
If $\pi < \pi_0$ for 5 consecutive iterations, 
$\lambda_1$ is decreased by a factor of 1.2.
The minimization of \eqref{objective}
is terminated when 
$\lambda_1$ does not increase for
25 consecutive iterations while
$\pi$ is maintained larger than $\pi_0$, indicating no further improvement of embedded consensus can be achieved.\\
\textbf{Inversion Settings:}
For trigger estimation, we utilized the same small clean dataset. Using the detected target class in \eqref{objective}, 
we report results for $\lambda_1$ and $\lambda_2$ fixed at $0.00001$ and $0.5$, respectively. We experimentally found that performance was not sensitive to these specific choices, see Appendix \ref{sec:inv-different-lambda}.
The minimization of \eqref{objective}
is terminated when the average perturbation norm $\|\Bar{\delta}\|$ does not decrease for 25 consecutive iterations while
$\pi$ is maintained larger than $\pi_0$.


\subsection{Backdoor Detection Performance}\label{sec:backdoor-detection-experiment}
We compared CEPA with seven state-of-the-art post-training detection methods: NC \cite{NC}, TABOR \cite{TABOR}, ABS \cite{ABS}, PT-RED \cite{TNNLS}, META \cite{meta_unsup}, TND \cite{DataLimited} and MMBD \cite{backdoor-universal}. 
\begin{table*}
	\begin{center}
			\begin{tabular}{cccccccc}
				\toprule
				   & $N_{\rm img}$    &  clean    &   chessboard    &  1-pixel     & BadNet    & unicolor   & blend\\
				\midrule
				NC\cite{NC}   & 10    & 56.7$\pm$49.6   &    100.0$\pm$0.0   & 36.7$\pm$48.2   &      96.7 $\pm$ 18.0   &  76.7$\pm$42.3   & 33.3$\pm$47.1\\
				TABOR\cite{TABOR}   & 10   & 70.0$\pm$45.8      &    93.3$\pm$24.9   &    93.3$\pm$24.9   &  73.3$\pm$44.2   &  66.7$\pm$47.1   &    86.7$\pm$34.0\\
				ABS\cite{ABS}   & 1    &   93.3$\pm$24.9   &  66.7$\pm$47.1   &     80.0$\pm$40.0   &      96.7$\pm$18.0   &  46.7$\pm$49.9   &       90.0$\pm$30.0\\
		META\cite{meta_unsup}   & 10k   &  73.3$\pm$44.2   &  60.0$\pm$49.0   & 6.7$\pm$24.9   &         83.3$\pm$37.3   &  30.0$\pm$45.8      &    86.7$\pm$34.0\\ 
    				TND\cite{DataLimited}     & 5    & 50.0$\pm$50.0    &  26.7$\pm$44.2    &     83.3$\pm$37.3    & 33.3$\pm$47.1    &  3.3$\pm$18.0       & 56.7$\pm$49.6\\
				PT-RED\cite{TNNLS}    & 100   & 76.7$\pm$42.3   &     96.7$\pm$18.0     &    100.0$\pm$0.0   &  13.3$\pm$34.0   & 23.3$\pm$42.3   &  63.3$\pm$48.2\\
    				MMBD\cite{backdoor-universal}    &  0    &      86.7$\pm$34.0    &    90.0$\pm$30.0      &      100.0$\pm$0.0    &           100.0$\pm$0.0    &        100.0$\pm$0.0   &         100.0$\pm$0.0\\
            CEPA $\|\Bar{\delta}\|$,$~ \ell 9$ & 10 & 90.0$\pm$30.0 & 0.0$\pm$0.0 & 73.3$\pm$44.2 & 0.0$\pm$0.0 & 53.3$\pm$49.9 & 16.7$\pm$37.3 \\
            CEPA $\|\mu\|$,$~ \ell 9$ & 10 & 100.0$\pm$0.0 & 93.3$\pm$24.9 & 46.7$\pm$49.9 & 96.7$\pm$18.0 & 13.3$\pm$34.0 & 76.7$\pm$42.3 \\
             CEPA $\frac{\sigma}{\|\mu\|}$,$~ \ell 9$ & 10 & 100.0$\pm$0.0 & 6.7$\pm$24.9 & 70.0$\pm$45.8 & 73.3$\pm$44.2 & 13.3$\pm$34.0 & 53.3$\pm$49.9 \\
             CEPA ens., $~ \ell 9$ & 10 & 90.0$\pm$30.0 & 96.7$\pm$18.0 & 80.0$\pm$40.0 & 96.7$\pm$18.0 & 66.7$\pm$47.1 & 93.3$\pm$24.9 \\
			\bottomrule
    \end{tabular}
	\caption{Detection accuracies ($\pm$ one standard deviation) on ResNet-18 for CIFAR-10 based on 30 clean models (for false-positive performance) and 30 attacked models for each of 5 different backdoors: additive global chessboard \cite{TNNLS}, additive one-pixel \cite{SS} \cite{AC}, BadNet patch \cite{BadNet}, unicolor patch, and blended patch \cite{NC}.  CEPA results are reported on model layer 9, based on $\|\Bar{\delta}\|, \|\mu\|$ and
 $\frac{\sigma}{\|\mu\|}$ respectively and collectively (ens.). Accuracy on clean models equals one minus the false-positive rate.}
		\label{tab:detection_accuracy}
	\end{center}
\end{table*}

Table \ref{tab:detection_accuracy} 
shows detection performance for the detectors on ResNet-18/CIFAR-10, leveraging the ensemble experimental results from \cite{backdoor-universal}. The ensembles consist of 30 clean and 30 attacked models for each attack type. The number of clean
images per class is reported as $N_{img}$.
A detection is considered accurate when the attack is detected and the correct target class is inferred. 
The last four rows demonstrate the detection results from CEPA's metrics: 1) $\|\Bar{\delta}\|$, 2) $\|\mu\|$, 3) $\frac{\sigma}{\|\mu\|}$, and `ens.' (ensemble), where a union rule on the individual metrics is applied.
Also, CEPA ens. true detection rate for the \textbf{WaNet} attack on PreAct-ResNet-18/CIFAR-10 is \textbf{7/10}, and for comparison MMBD's is 10/10.  
CEPA ens. on the clean PreAct-ResNet-18/CIFAR-10 model  achieves 9/10 accuracy.
 
For CEPA, we chose the 9th layer of the ResNet-18 model as the embedded feature layer as it exhibits the best performance. Again, recall that CEPA can be applied to {\it every} layer, with detection then based on the layer with the most extreme detection statistic. Thus, CEPA performance in the table could potentially improve when more layers are considered. See the next subsection.

A CEPA detection is made when one or more of the three metrics follows the assumed anomaly direction and its MAD exceeds the detection threshold. Conversely, we determine the class as clean if all three metrics do not show MAD above the threshold in the anomaly direction. For $\|\Bar{\delta}\|$ and $\frac{\sigma}{\|\mu\|}$, 
the MAD threshold of 2 is used, corresponding to two standard deviations beyond the median. For $\|\mu\|$, we used a MAD threshold of $3$, as it exhibits more extreme outliers. The MAD thresholds are kept fixed in all our experiments.

As discussed in Appendix \ref{sec:prior}, existing post-training detectors are often tailored for specific backdoor pattern types and incorporation mechanisms. NC gives strong results for patch replacement (BadNet), but poor results for the local additive pattern (1-pixel) and blend attacks. 
Similarly, ABS and META gives poor results for unicolor and chessboard, and META also fails for the unicolor attack.
Conversely, PT-RED is optimized for additive patterns (chessboard, 1-pixel) and struggles with other types like BadNet and blend. TABOR and TND perform less competitively because they make additional assumptions about the shape or color of the backdoor pattern. By contrast, CEPA outperforms the other detectors across different attacks in general, except for MMBD. CEPA demonstrates overall good and robust (across attack types) detection accuracy with a low false detection rate for clean models. CEPA's universal detection power is further demonstrated by its good performance on WaNet, a more recent, advanced backdoor incorporation mechanism.

One can observe that all three CEPA metrics are essential for robust detection performance. $\|\mu\|$ effectively identifies chessboard, BadNet, and blend attacks, while $\frac{\sigma}{\|\mu\|}$ provides complementary detection for 1-pixel, BadNet, and blend. Conversely, $\|\Bar{\delta}\|$ demonstrates strong performance in detecting 1-pixel and unicolor patch attacks. It is important to mention that $\|\mu\|$ and $\frac{\sigma}{\|\mu\|}$ exhibit zero false positive rates, underscoring their exceptional reliability on clean models.

Finally, note that CEPA is competitive with the very best overall method -- MMBD -- while achieving lower false positive detections.  Moreover, unlike MMBD, CEPA reverse-engineers the backdoor pattern as we will see in the sequel.

\subsection{Detailed Illustration on CEPA Detection Performance}
To further illustrate typical CEPA detection performance, we provide detailed analysis for single typical ResNet18/CIFAR-10 models under 
chessboard, unicolor patch, Hello Kitty blend, and WaNet attacks, along with a clean model for comparison.
All backdoors attacks and training settings are consistent with section \ref{sec:experiment-setup} except that here we fixed the poison rate at 1.8\% and the ground truth target class as class 9. Layers 5, 9 and 13 were selected as embedded feature layer candidates.

Table \ref{tab:MAD-10} shows CEPA's largest MAD scores, measured in the correct direction, for the three detection metrics, $\|\Bar{\delta}\|$, $\|\mu\|$, and $\frac{\sigma}{\|\mu\|}$, for layers 5, 9, and 13,
along with the class that achieves that score (shown in the parenthesis). Note that the true target class, class 9, achieves more extreme MAD scores than the other (non-target) classes.

Both layers 5 and 9 exhibit strong detection performance while use of layer 13 (closer to the output layer) is much less effective.
WaNet is only detected at
layer 5 and 9 using $\|\mu\|$.
Also, the chessboard pattern is detected  by both $\|\mu\|$ and $\frac{\sigma}{\|\mu\|}$ at layer 9 but not at layer 5. For layer 5, there is a false positive detection using \(\|\Bar{\delta}\|\) on the clean model, but no false alarms using layer 9.
In general, layer 9 exhibits the best performance across attacks, without any false positives on clean models.
That is, it appears that the consensus embedded pattern expresses most convincingly in layer 9, compared with layers too close to the input (layer 5) and too close to the decision-making layer (layer 13).
Again, CEPA can be applied using a plurality of layers, with the detection decision based on the layer and statistic exhibiting the most pronounced MAD score.

\begin{figure*}[t]
     \centering
     \begin{subfigure}[b]{0.18\textwidth}
         \centering
         \includegraphics[width=\textwidth]{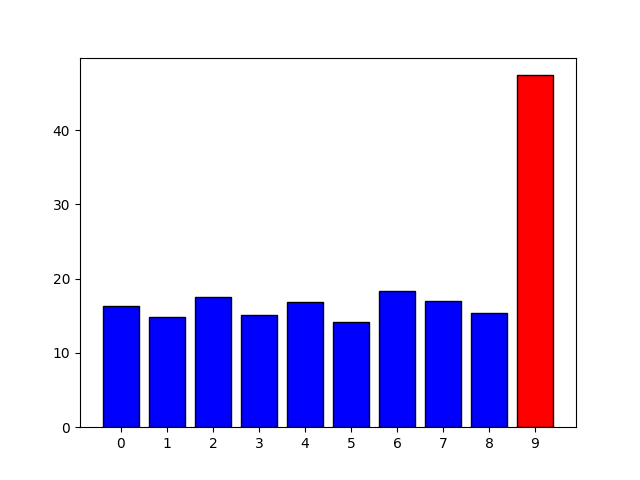}
         \caption{BadNet}
     \end{subfigure}
     \begin{subfigure}[b]{0.18\textwidth}
         \centering
         \includegraphics[width=\textwidth]{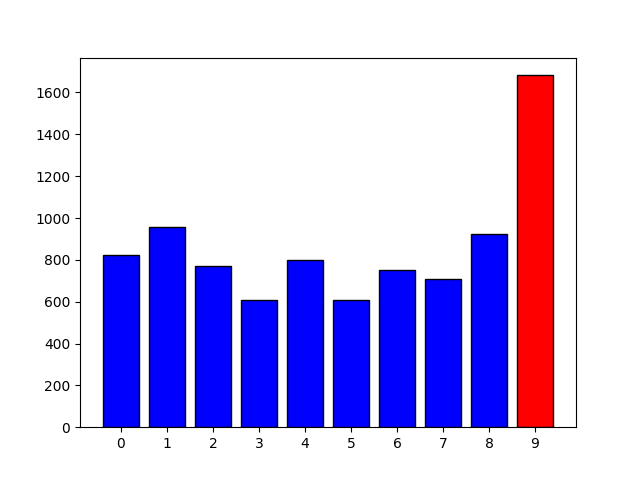}
         \caption{WaNet}
     \end{subfigure}
     \begin{subfigure}[b]{0.18\textwidth}
         \centering
         \includegraphics[width=\textwidth]{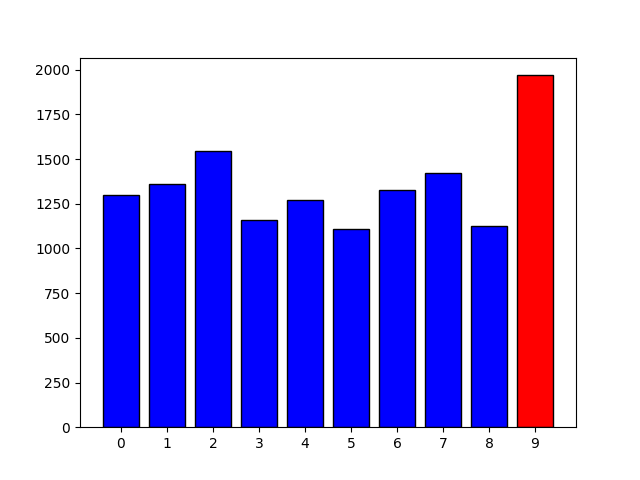}
         \caption{chessboard}
     \end{subfigure}
     \begin{subfigure}[b]{0.18\textwidth}
         \centering
         \includegraphics[width=\textwidth]{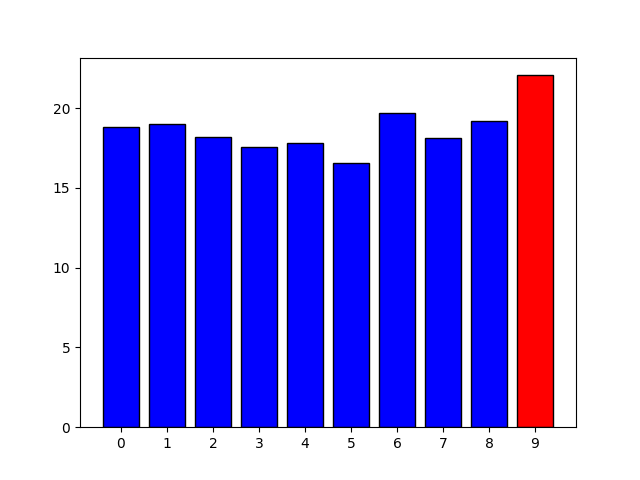}
         \caption{blend}
     \end{subfigure}
     \begin{subfigure}[b]{0.18\textwidth}
         \centering
         \includegraphics[width=\textwidth]{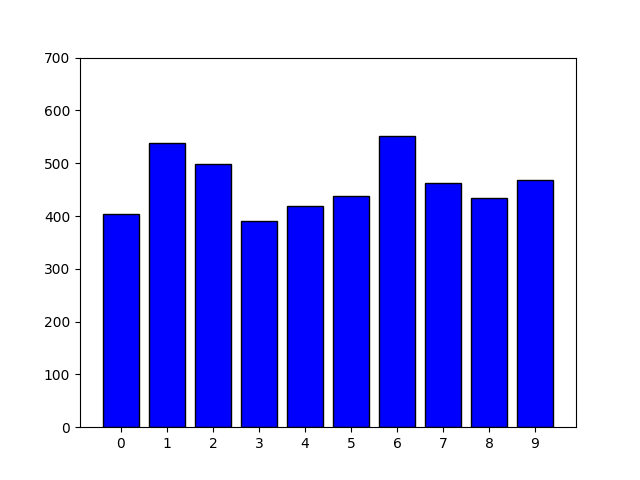}
         \caption{clean}
     \end{subfigure}
        \caption{Layer 9, $\|\mu\|$ values by class, for typical poisoned and clean
        models with true target class in red.}
        \label{fig:layer9-norm}
        \vspace{-0.1in}
\end{figure*}

Figure \ref{fig:layer9-norm} shows a bar chart of the detection metric $\|\mu\|$ for these five models on layer 9.
From the bar chart (one bar for each {\em putative} target class), one can see that, our method strongly indicates the true target class (9, red bar) on poisoned models, i.e., the bars for the true target class are clear visual outliers compared to those for the non-target classes (0-8), but does not indicate any significant outliers in the case of a clean model (there are no false positive class detections). Additional bar charts associated with Table \ref{tab:MAD-10} can be found in Figure \ref{fig:layer9-consensus}
and Section  \ref{sec:barcharts_for_layer5_13} of the Appendix.

Finally, we note that CEPA is insensitive to the threshold on the misclassification rate, $\pi_0$. For $\pi_0=0.5, 0.8,$ and $ 0.9$, CEPA demonstrates stably strong detection results. See Appendix \ref{sec:ablation} for more details.


\begin{table}
    \centering
    \begin{tabular}{c|ccccc}
    \toprule
     layer, crit. & BN & WN & CB & Bl. & Clean  \\ 
     \hline
     5,  $\|\Bar{\delta}\|$   
     & 1.22(5) & 1.01(3) & 1.95(5) & 2.11(9) & 2.21(9) \\ 
     5,  $\|\mu\|$   
     & 20.95(9) & 9.96(9) & 2.70(9) & 4.32(9) & 1.21(9) \\ 
     5,  $\frac{\sigma}{\|\mu\|}$   
     & 4.88(9) & 1.54(9) & 2.10(2) & 3.54(9) & 1.24(2) \\ 
     9,  $\|\textstyle\Bar{\delta}\|$   
     & 1.90(5) & 1.24(3) & 1.29(3) & 1.63(4) & 0.71(0) \\ 
     9,  $\|\textstyle\mu\|$   
     & 15.77(9) & 5.59(9) & 3.42(9) & 3.41(9) & 1.76(6) \\ 
     9,  $\frac{\sigma}{\|\mu\|}$   
     & 5.82(9) & 1.08(9) & 2.06(9) & 2.72(9) & 1.47(1) \\
     13,  $\|\Bar{\delta}\|$   
     & 5.11(9) & 2.07(3) & 0.94(3) & 1.23(3) & 1.30(6) \\ 
     13, $\|\mu\|$   
     & 2.02(8) & 2.77(9) & 2.64(1) & 0.94(1) & 1.87(7) \\ 
     13, $\frac{\sigma}{\|\mu\|}$   
     & 1.24(2) & 0.86(7) & 1.22(7) & 0.97(1) & 1.09(5) \\ 
    \bottomrule
    \end{tabular}
    \caption{CEPA MAD scores for different attacks against ResNet-18/CIFAR-10 and for the Clean (unpoisoned) model, on layers 5, 9, and 13 for the 3 detection metrics. The score represents the largest MAD value among all potential target classes. Along with each score, the class achieving that score is indicated in the parenthesis. The attacks are
    BadNet (BN), WaNet (WN), chessboard (CB), and Blend (Bl.).}
    \label{tab:MAD-10}
    \vspace{-0.1in}
\end{table}

\subsection{Backdoor Inversion Performance}
\label{sec:backdoor-inv-performance}
Recall from section \ref{sec:backdoor-inversion} that CEPA reverse-engineering involves optimization of the same objective function as for detection.  Following this optimization, the estimated 
${\delta}_x, \forall x \in \Dcal_s$ are our (sample-wise) estimates of the backdoor pattern in the input space.  Here we investigate how revealing this approach is about the ground-truth backdoor pattern.
%
%

\begin{figure}[!ht]
    \centering
    \begin{subfigure}[b]{0.28\columnwidth}
         \includegraphics[width=\textwidth]{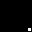}
         \caption{}
         \label{fig:badnet-gt}
     \end{subfigure}
    \begin{subfigure}[b]{0.28\columnwidth}
        \includegraphics[width=\textwidth]{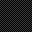}
         \caption{}
         \label{fig:chessboard-gt}
     \end{subfigure}
     \begin{subfigure}[b]{0.28\columnwidth}
        \includegraphics[width=\textwidth]{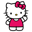}
         \caption{}
         \label{fig:blend-gt}
     \end{subfigure}
     \\
    \begin{subfigure}[b]{0.28\columnwidth}
         \includegraphics[width=\textwidth]{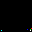}
         \caption{}
         \label{fig:badnet-inv}
     \end{subfigure}      
    \begin{subfigure}[b]{0.28\columnwidth}
        \includegraphics[width=\textwidth]{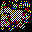}
         \caption{}
         \label{fig:chessboard-inv}
     \end{subfigure}
     \begin{subfigure}[b]{0.28\columnwidth}
        \includegraphics[width=\textwidth]{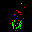}
         \caption{}
         \label{fig:blend-inv}
     \end{subfigure}
         \caption{For the BadNet attack: (a) the ground truth perturbation, and (d) the average estimated perturbation $|\mathcal{D}|^{-1}\sum_{x\in \mathcal{D}}\hat{\delta}_x$.
         For the chessboard attack: (b) the ground truth perturbation (with contrast boosted for better visualization), and  (e) a typical $\delta$. For the blended attack: (c) the ground truth perturbation, and (f) $|\mathcal{D}|^{-1}\sum_{x\in \mathcal{D}}\hat{\delta}_x$. }
         \label{fig:visual_BadNet}
         \vspace{-0.1in}
         \end{figure}

For ResNet-18/CIFAR-10,
Figure \ref{fig:visual_BadNet} visualizes CEPA backdoor-inversion
results for BadNet patch, additive chessboard, and blend attacks with $\lambda_1=0.00001, \lambda_2=0.5$.
It is important to note that, in CEPA's loss function, triggers are (sample-wise) additively applied to images.  The resulting, estimated sample-wise $\delta_x$'s
may in fact have non-zero values at all pixel locations, albeit {\it small} values at pixel locations that are {\it not} part of the ground-truth backdoor pattern's support, for localized patch replacement attacks.  Indeed,
Figure \ref{fig:visual_BadNet}(a) shows the ground truth BadNet trigger (white color patch in lower right of the image), and
(d) shows the average estimated perturbation
$\Bar{\delta}$ over all clean samples from the source classes, $\mathcal{D}$.
While there is discrepancy in appearance between the ground truth and the estimated trigger, the estimated trigger is localized and is in the same location as the ground-truth trigger.
Moreover, the ``artifact'' in the lower left of Figure \ref{fig:visual_BadNet}(d) is likely due to use of data augmentation by horizontal flipping during the model's training.
Figure \ref{fig:visual_BadNet}(b) shows the ground truth trigger pattern for the additive chessboard attack and (e) shows a typical CEPA-estimated pattern $\delta_x$, where one can clearly see a portion of the ground-truth backdoor pattern. Figure \ref{fig:visual_BadNet}(c) presents the ground truth Hello Kitty pattern for the global blend attack and (f) shows the average estimated perturbation $\Bar{\delta}$. Without knowing the blend ratio for the backdoor pattern, CEPA managed to estimate the shape and position of some color patches in the true pattern, e.g. the pink skirt is reflected as the green pixels in a similar location; Kitty's hands and feet are estimated by the red pixels with waving and standing posture. We also found that accurate inversion is
not very sensitive to the choice of the $\lambda$s; see Appendix \ref{sec:inv-different-lambda}.

Although it is difficult to make general claims about our accuracy in revealing the backdoor incorporation mechanism, the patterns we reverse-engineer offer a form of "explainable AI". For instance, in the case of patch attacks, the reverse-engineered pattern is spatially localized in the corners of the image rather than the foreground. This spatial location raises suspicion of a backdoor since legitimate salient features are typically found in the image foreground, not the background or corners. Similarly, for chessboard and WaNet, the reverse-engineered patterns reveal global changes affecting class decisions, which may be indicative of a global backdoor attack pattern.

\begin{table}
    \centering
    \begin{tabular}{l|cccc}
    \toprule
    ASR & BadNet & blend  & WaNet & chessboard \\
    \hline
    GT  & 95.73\% & 91.53\%  & 97.31\%  & 96.85\% \\ 
    \eqref{objective}     & 92.22\%   & 93.33\%    & 96.67\%   & 91.11\% \\ 
    \bottomrule
    \end{tabular}
        \caption{Attack success rates when: i) the ground-truth (GT) backdoor pattern is incorporated into the evaluation samples;  
and
    ii) perturbations $\bar{\delta}$ obtained from \eqref{objective} are incorporated into the samples of the small clean dataset $\Dcal$ use for defense. 
    }
    \label{tab:ASR}
    \vspace{-0.2in}
\end{table}

To further illustrate effectiveness of the inverted triggers, we evaluate attack success rates (ASRs) on the test set embedded with the average estimated perturbation $\Bar{\delta}$, shown in Table \ref{tab:ASR}. Note that the CEPA-estimated trigger achieves high ASRs, comparable to those achieved using the ground-truth trigger.
While UNICORN \cite{UNICORN23} reports 
higher ASRs (see their Table 1),
their poisoning fraction is $2.8$ times higher (5\%, see their Section A.5, 
compared to our 1.8\%), which may make the backdoor pattern more conspicuous and more easily detected. 
Note, though, that a poisoning rate of just 1.8\% 
is in fact sufficient to achieve a very high attack success rate for the adversary.

As discussed in Appendix \ref{sec:prior}, if UNICORN's feature maps are trained using the
(possibly poisoned) training set of the defended model, then it is
not truly a post-training defense.
Again, the post-training scenario is important considering
that, e.g., AIs are sometimes built from foundation models acquired
from other parties or are produced by other parties and 
embedded in larger systems which are then acquired (with potentially no ``data rights'' access to the training set, for the party who acquires the model).

Additionally, CEPA
\begin{itemize}[leftmargin=5mm]
\item truly possesses only a single hyperparameter (the misclassification fraction threshold, used to terminate the CEPA optimization), as the internal layer $f(\cdot)$ and the $\lambda$s can be automatically chosen;
\item does not involve a bound on the ``size" of the backdoor pattern for detection (unlike UNICORN's ``mask size constraint") -- indeed, it detects the chessboard pattern, which has global pixel support;  
\item demonstrates that auxiliary feature maps (as used by UNICORN) are not necessary and that the embedded features of the defended model suffice for achieving accurate agnostic backdoor detection and reverse engineering.
\end{itemize}
\subsection{VGG-16 Model for the CIFAR-100 Dataset}

\begin{figure}[!ht]
    \centering
    \begin{tabular}{cc}
    \includegraphics[width=0.96\columnwidth]{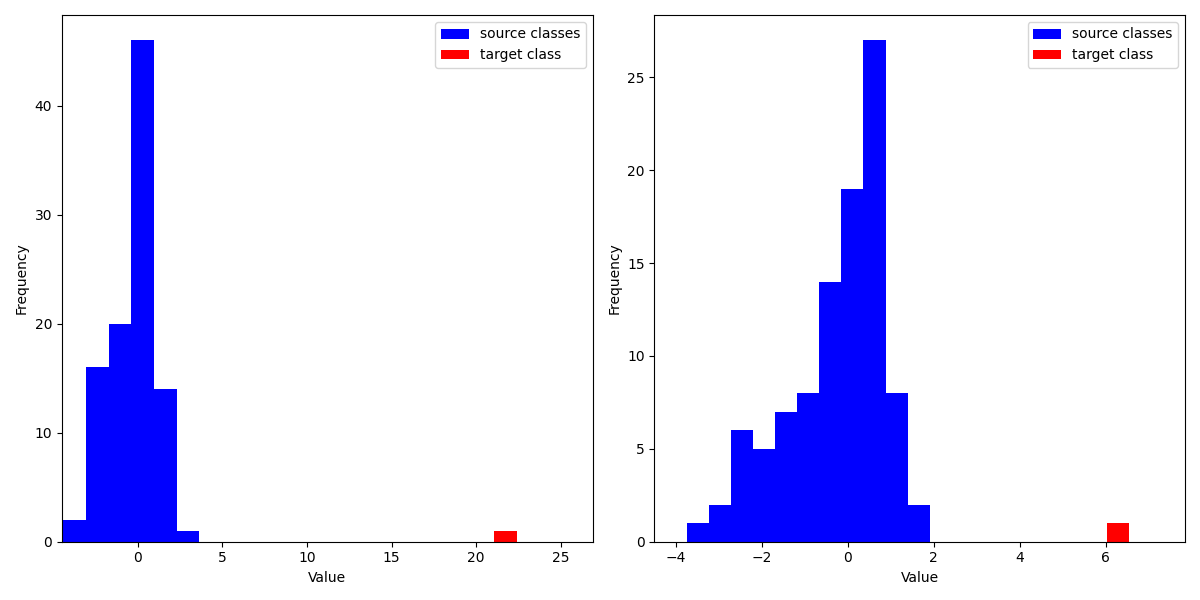}
    \end{tabular}
    \vspace{-0.2in}
    \caption{Histogram of MAD anomaly scores for CEPA against 
    BadNet on a typical poisoned VGG-16 model for CIFAR-100 calculated at layer 8 using $\|\mu\|$ (left) and  $\frac{\sigma}{\|\mu\|}$ (right).}
    \vspace{-0.2in}
    \label{fig:vgg_MAD}
\end{figure}

CEPA has similarly good detection results on the $3^{\rm th}$ and $8^{\rm th}$ layers of VGG-16 trained on CIFAR-100. MAD scores on layer 8 are shown in Figure \ref{fig:vgg_MAD} against BadNet, where the ground truth backdoor target class 0 is marked in red. More details can be found in Appendix \ref{sec:CIFAR-100}.

\subsection{Adaptive Attacks}

CEPA performs well in the presence of the adaptive attack considered in \cite{backdoor-universal}.
See Appendix \ref{sec:adaptive} for results of an adaptive attack on ResNet-18/CIFAR-10.

\subsection{Time Complexity}
Our experiments utilize a NVIDIA RTX 3090 GPU with 24GB of memory. CEPA requires approximately 1.5 hours to perform detection on a ResNet-18 model trained on CIFAR-10, using 10 clean images per class (90 images total). On a VGG-16 model trained on CIFAR-100 using 2 clean images per class (198 images total), CEPA takes approximately 11 hours. For comparison, training a Resnet-18 on CIFAR-10 for 60 epochs takes approximately 40 minutes, and training a VGG-16 on CIFAR-100 for 100 epochs takes 3 hours. Note that the detection time complexity is mainly due to the grid search of $\lambda_1$.
For backdoor reverse engineering without grid search, CEPA only requires about 10 minutes.
\section{Summary and Discussion}

This paper proposes
the CEPA post-training defense
against backdoor poisoning in deep neural networks.
CEPA detects whether the model
is backdoor poisoned and, if so, identifies 
the target class.  CEPA can also  
estimate the backdoor trigger pattern for a given sample.
CEPA
is agnostic to the manner of incorporation of the
backdoor pattern.
CEPA 
requires few clean samples; it requires effectively only a single hyperparameter; it relies only on embedded feature activations
(not on assumed clean, auxiliary feature maps, as for \cite{UNICORN23}); 
and is an effective detector. 
CEPA's 
$\lambda$s can be dynamically adjusted during the optimization process, and multiple choices for the embedded feature layer
can be simultaneously assessed without very large computational overhead.
Finally, 
backdoor inversion results show CEPA reveals key features of backdoor trigger patterns.


\clearpage


\clearpage
\appendix
\section{Appendix}

\subsection{Prior work on agnostic backdoor defenses}\label{sec:prior}

Though not all post-training backdoor detection
defenses are reverse-engineering based, 
e.g., \cite{KLiu18,UIUC-TrojAI,MMBD}, 
several reverse-engineering defenses 
have been proposed, e.g., \cite{NC,ABS,TNNLS},
all relying upon a
small, clean (unpoisoned, correctly labeled)
dataset (which may be disjoint from
the training set of the DNN under consideration).

\cite{TNNLS} leverages a small, clean 
dataset 
to reverse-engineer putative backdoor patterns.  This is achieved by seeking the smallest common additive perturbation that induces most patterns from one class (a putative source class) to be misclassified to another class (a putative target class).
The common additive perturbation is {\it either} estimated at the input layer or in an internal (embedded) layer of the DNN.
Working on an embedded layer is motivated by cases where the input space is discrete-valued (hence unsuitable for gradient-based estimation of perturbations)
or when the embedded feature dimension is much smaller than the
input dimension, e.g. for a Video Transformer model.  A backdoor pattern estimated in an 
embedded space can be mapped back to the input space for
purposes of interpretation using methods such as Grad-CAM
\cite{integrated-gradient}.
It was also demonstrated in \cite{TNNLS} (but in a quite limited way) that estimating an {\it additive} backdoor pattern
in an embedded feature space of the DNN
is {\it somewhat} agnostic to the method of backdoor incorporation --
a multiplicatively incorporated backdoor was detected working from an internal 
layer of the DNN.  
However, other non-additive backdoor-incorporation mechanisms were not investigated in \cite{TNNLS}. 

UNICORN and FeatureRE \cite{ZWang22,UNICORN23} are 
both reverse-engineering detectors
that attempts to be agnostic to the backdoor-incorporation mechanism. Their objective are designed based on the observation that neuron activation values representing the backdoor behavior are orthogonal to others, and they try to guarantee the orthogonality on a small set of neurons that are most activated by the backdoor.
Considering only all-to-one backdoor attacks, an (unpoisoned) UNet feature map (neural networks) is assumed available
for each (putative target) class 
in order to produce, from the model's input space,
a feature space within which the backdoor pattern
(hopefully) expresses either as a (BadNet) patch \cite{BadNet} 
or as a (Neural Cleanse - NC) blended patch \cite{NC}, see equation (4) of 
\cite{UNICORN23}.

Thus, UNICORN's design either requires suitable
unpoisoned feature maps to (somehow) be available or requires substantial (particularly when the number of classes is large) unpoisoned data and computation resources to learn them.
Such learning must be performed \textbf{before} the backdoor 
reverse-engineering process (NC) can commence.
Moreover, 
to account for possible one-to-one attacks, 
these methods 
would need to train two UNets for every putative (source,target) class pair, which 
would entail huge computation
even for a small number of classes (e.g.,
for 10 classes there are 90 class pairs).
Finally, note that given such clean feature maps, one may be able to 
append a fully connected (linear) layer to them and train
this layer with a small amount of
clean data (whose availability is commonly assumed by backdoor detection defenses), resulting in an accurate classifier.
That is, the availability of such feature maps may preclude the need for 
backdoor defenses.


Note also that, in principle, one could form an ensemble of reverse-engineering based detectors, with each such detector assuming a particular method of backdoor incorporation.  However, for ``digital'' attacks (as discussed before), there are effectively an {\it uncountably infinite number} of possible backdoor incorporation mechanisms.  Even covering a substantial number of the possible mechanisms would entail an ensemble system of very high complexity (both with respect to detector design as well as detection inference complexity), while still failing to cover myriad possible mechanisms.

A backdoor-agnostic approach that is not reverse-engineering based 
is \cite{MMBD}.  This work hypothesizes and experimentally verifies that backdoor-poisoned DNNs {\it overfit} to the backdoor pattern.  Such overfitting is necessary in order for the backdoor's features to overcome the ``normal'' source-class discriminative features, which would otherwise ensure classification to the (true) source class of the sample.   Overfitting to the backdoor pattern implies that the backdoor pattern triggers large internal-layer DNN signals, which are particularly enabled by (positively) unbounded ReLU activations.  \cite{MMBD} also demonstrated that optimizing saturation (clipping) levels on the ReLUs is effective at eliminating the learned backdoor mapping from the trained model.  
However, the approach in \cite{MMBD} does not reverse-engineer the backdoor pattern and also has no way to  exploit available clean data (to improve the detection performance).

Another important aspect is that, whereas for some backdoor attacks the backdoor pattern is a {\it static} pattern (e.g., the same pattern {\it added} to every poisoned training image) for other attacks, the backdoor pattern is effectively {\it sample-dependent}.  The backdoor pattern induces sample-dependent
{\it changes} both in the case of the warping attack \cite{WaNet} {\it as well as} in the case of patch replacement attacks (which, e.g. for digital images, involve replacing the pixel intensities in the backdoor patch region by the intensities of the backdoor patch).

The backdoor detector CEPA 
proposed herein has several key characteristics, exploiting some ideas from prior works:
i) it can be used to estimate (invert, reverse engineer) the backdoor pattern like \cite{NC,TNNLS,UNICORN23} and unlike \cite{MMBD};
ii) it attempts to identify an  \textbf{embedded} (internal layer) additive consensus
perturbation that corresponds to the
backdoor, like \cite{TNNLS}\footnote{This hypothesis of additivity in the embedded feature space is motivated by the observation that components of embedded feature-maps of a DNN are typically weighted and {\em superposed} in subsequent processing by, e.g., convolutional or linear layers.};
iii) it exploits the observation from \cite{MMBD} that 
the trained model overfits to the backdoor pattern, with this
overfitting achieved by the backdoor pattern eliciting large internal-layer signals in the DNN;
iv) like \cite{I-BAU}, it estimates {\it sample-specific}, backdoor mechanism-agnostic perturbations (thus enabling detection of backdoor attacks that involve sample-dependent changes such
as patch replacement \cite{BadNet} or warping \cite{WaNet}).
While the backdoor pattern may cause sample-dependent changes (i.e., at the input to the DNN), 
these sample-dependent changes may be inducing a very similar (i.e., $\sim$ a {\it common}) pattern in some internal layer of the DNN,
i.e., a pattern that the downstream network can learn to recognize (and which induces the model to misclassify to the attacker's target class).  Thus, we seek a {\it common} (consensus) backdoor pattern, not in the input sample, but in internal layer features of
the DNN that are induced by input samples.  Moreover, 
consistent with the backdoor overfitting phenomenon \cite{MMBD}, this common backdoor pattern should express as a {\it large} internal-layer perturbation (even though the {\it input} perturbation may be small
in magnitude 
in order to be inconspicuous).

Note that, unlike UNICORN \cite{ZWang22,UNICORN23} that postulate that the backdoor is activating on only a small portion of the embedded neurons, we look at the general behavior of embedded layer as a whole.


\subsection{Experiment Setup Details} \label{sec:setup_details}

\subsubsection{Details on Backdoor Triggers}
\begin{figure}[h!]
    \vspace{-0.15in}
    \captionsetup[subfigure]{justification=centering}
    \centering
    \begin{subfigure}[t]{0.12\textwidth}
        \includegraphics[width=\textwidth]{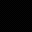}
    \end{subfigure}
    \begin{subfigure}[t]{0.12\textwidth}
        \includegraphics[width=\textwidth]{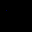}
    \end{subfigure}
    \begin{subfigure}[t]{0.12\textwidth}
        \includegraphics[width=\textwidth]{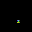}
    \end{subfigure}
    \begin{subfigure}[t]{0.12\textwidth}
        \includegraphics[width=\textwidth]{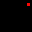}
    \end{subfigure}
    \begin{subfigure}[t]{0.12\textwidth}
        \includegraphics[width=\textwidth]{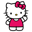}
    \end{subfigure}
    \begin{subfigure}[t]{0.12\textwidth}
        \includegraphics[width=\textwidth]{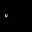}
    \end{subfigure}
    \begin{subfigure}[t]{0.12\textwidth}
        \includegraphics[width=\textwidth]{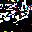}
    \end{subfigure}
    \par\vspace{0pt}
    \begin{subfigure}[t]{0.12\textwidth}
        \includegraphics[width=\textwidth]{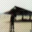}
        \caption*{chessboard}
    \end{subfigure}
    \begin{subfigure}[t]{0.12\textwidth}
        \includegraphics[width=\textwidth]{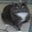}
        \caption*{1-pixel}
    \end{subfigure}
    \begin{subfigure}[t]{0.12\textwidth}
        \includegraphics[width=\textwidth]{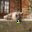}
        \caption*{BadNet}
    \end{subfigure}
    \begin{subfigure}[t]{0.12\textwidth}
        \includegraphics[width=\textwidth]{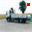}
        \caption*{unicolor}
    \end{subfigure}
    \begin{subfigure}[t]{0.12\textwidth}
        \includegraphics[width=\textwidth]{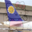}
        \caption*{global blend}
    \end{subfigure}
    \begin{subfigure}[t]{0.12\textwidth}
        \includegraphics[width=\textwidth]{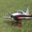}
        \caption*{local blend}
    \end{subfigure}
    \begin{subfigure}[t]{0.12\textwidth}
        \includegraphics[width=\textwidth]{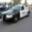}
        \caption*{WaNet}
    \end{subfigure}
    \caption{Example backdoor patterns (top) and corresponding poisoned images (bottom) used in our experiments.}
    \label{fig:BP_example}
\end{figure}

On CIFAR-10, we consider the following triggers with different backdoor embedding types including additive, patch replacement, blending and warping:
1) a global additive perturbation (with size 3/255) resembling a chessboard
\cite{TNNLS};
2) a 1-pixel additive backdoor pattern which perturbs a single, randomly selected pixel by 75/255 in all color channels, with this pixel fixed for all trigger images for a given attack \cite{SS} \cite{AC};
3) a $3\times3$ patch (BadNet) with a randomly selected location (fixed for all trigger images for a given attack), used in \cite{BadNets};
4) a $3\times3$ patch similar to BadNet but monochromatic (unicolor);
5) a $3\times3$ local random patch trigger with a blend ratio of $\alpha=0.2$, with a randomly selected and fixed location for each image in a given attack, used in \cite{Targeted-Backdoor};
6) a global Hello Kitty trigger with a blend ratio of $\alpha=0.15$ used by \cite{Targeted-Backdoor};
7) a warping-based trigger (WaNet) proposed by \cite{WaNet}.

Note that of these, BadNet can be physically implemented, while
some digital attacks (e.g., WaNet) are more involved to implement than others.
We mainly considered all-to-one attacks, with a random class chosen as the target class.

On CIFAR-100 we only considered the BadNet unicolor patch. 
Our CIFAR-100 experiment illustrates the scalability of CEPA to large-class domains.

For CIFAR-10, an ensemble of 30 attacked models was produced for attacks 1)-5), along with an ensemble of 30 clean models. 
The ensemble of clean classifiers was used to evaluate the false positive detection rate.
For attack 7) an ensemble of 10 attacked models and an ensemble of 10 clean models was produced.
For attack 6), only a single (illustrative) attack realization was produced.
For CIFAR-100, we created one attacked model for the BadNet attack and one clean model.

In Figure \ref{fig:BP_example}, we demonstrate the different backdoor trigger patterns and corresponding poisoned images, with the backdoor trigger incorporated. Note that for WaNet, the 'pattern' shown is the attacked image minus the original.

\subsubsection{Details on Attack Settings}
\begin{table}[h!]
    \centering
    \begin{tabular}{lccccc}
        \toprule
        & chessboard  & 1-pixel & BadNet & unicolor & blend \\
        \midrule
        PR(\%) & 4.0 & 2.0 & 1.0 & 1.0 & 2.0 \\
        ASR(\%) & 99.94 & 91.28 & 99.92 & 98.44 & 96.86 \\
        ACC(\%) & 91.12 & 91.12 & 91.59 & 91.36 & 96.48 \\
        \bottomrule
    \end{tabular}
    \caption{For each attack on ResNet-18/CIFAR-10, the poison rate (PR), average attack success rate (ASR) and ACC (clean test accuracy) over the ensemble of 30 attacked models.}
    \label{table:setup_ASR}
\end{table}

On ResNet-18/CIFAR-10, Table \ref{table:setup_ASR} lists the poison rate (PR) for each attack, the average attack success rate (ASR) as well as the clean test accuracy (ACC) over the 30 attacked models.
For WaNet's 10-model ensemble on PreActResNet-18, the PR is 10\% and the average ASR and ACC are 96.45\% and 91.02\%, respectively.

On VGG-16/CIFAR-100, for the BadNet model, the poison rate was 1.8\% and the ASR and ACC are 95.63\% and 65.34\%, respectively. 
On the clean model, the ACC was 66.17\%.

Table \ref{tab:training_setting} records the training settings for all attacked and clean models for CIFAR-10 and CIFAR-100.
\begin{table}[ht]
\centering
\begin{tabular}{ccc}
\toprule
& CIFAR-10 & CIFAR-100 \\ 
\hline
model & ResNet-18 & VGG-16 \\ 

optimizer & Adam & Adam \\ 

batch size & 128 & 128 \\ 

epochs & 60 & 100 \\ 

learning rate & 1e-3 & 1e-4 \\ 
  \bottomrule
\end{tabular}
\caption{Training settings for CIFAR-10 and CIFAR-100 datasets.}
\label{tab:training_setting}
\end{table}

\subsection{Layer 9 Bar Charts of CEPA detection metric $\frac{\sigma}{\|\mu\|}$ on ResNet-18 for CIFAR-10}  
\begin{figure*}[t]
     \centering
     \begin{subfigure}[b]{0.18\textwidth}
         \centering
         \includegraphics[width=\textwidth]{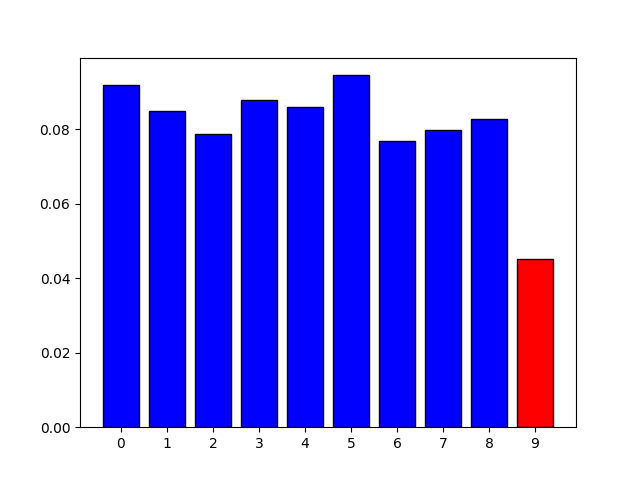}
         \caption{BadNet}
     \end{subfigure}
     \begin{subfigure}[b]{0.18\textwidth}
         \centering
         \includegraphics[width=\textwidth]{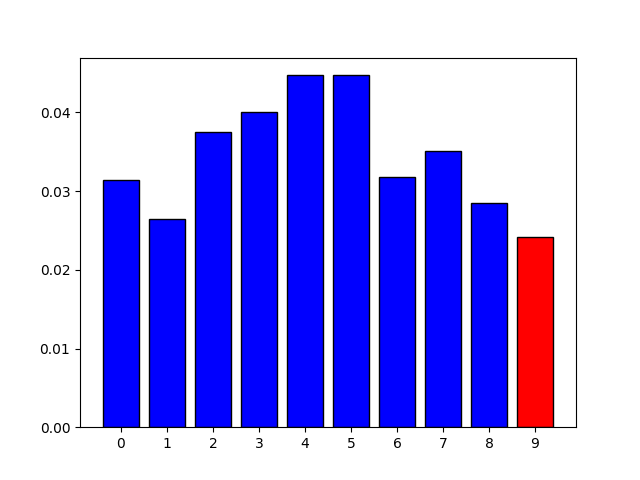}
         \caption{WaNet}
     \end{subfigure}
     \begin{subfigure}[b]{0.18\textwidth}
         \centering
         \includegraphics[width=\textwidth]{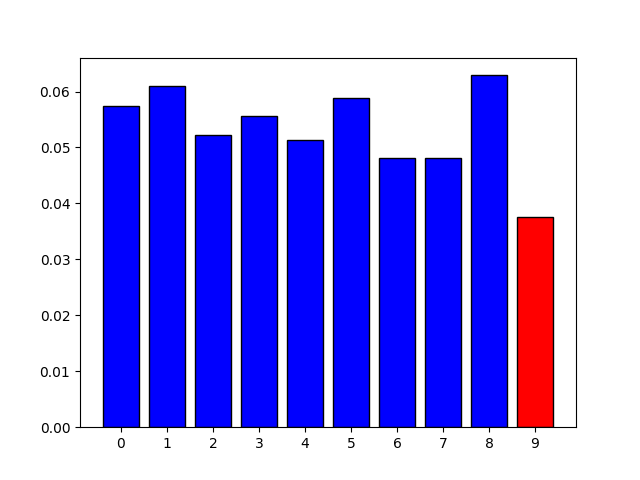}
         \caption{chessboard}
     \end{subfigure}
     \begin{subfigure}[b]{0.18\textwidth}
         \centering
         \includegraphics[width=\textwidth]{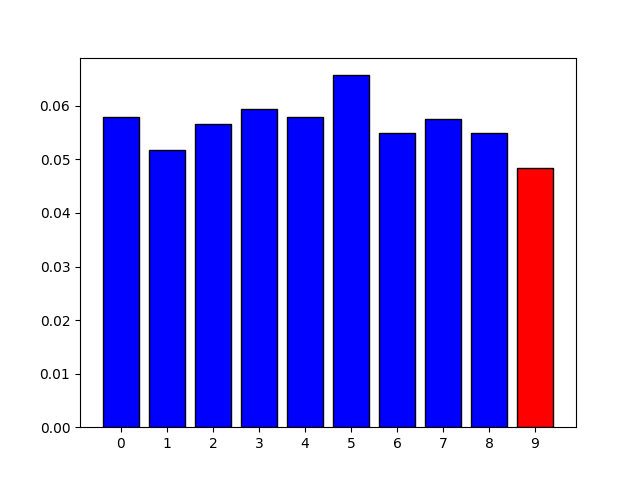}
         \caption{blend}
     \end{subfigure}
     \begin{subfigure}[b]{0.18\textwidth}
         \centering
         \includegraphics[width=\textwidth]{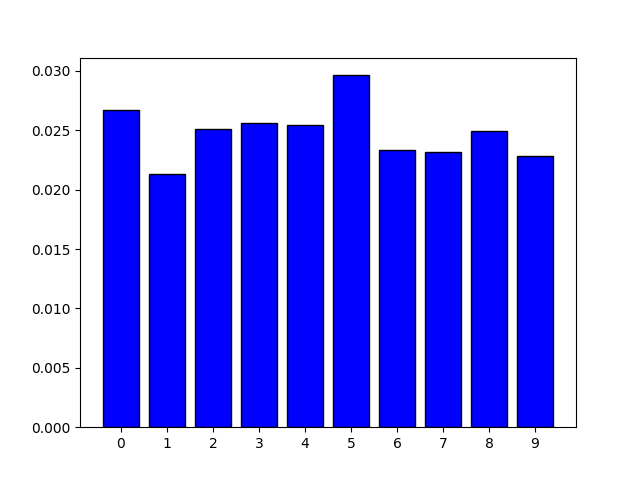}
         \caption{clean}
     \end{subfigure}
        \caption{Layer 9, $\sigma/\|\mu\|$ values by class, for typical poisoned and clean
        models, with the true target class in red.}
        \label{fig:layer9-consensus}
\end{figure*}

Figure \ref{fig:layer9-consensus} shows CEPA results on $\frac{\sigma}{\|\mu\|}$, for layer 9, for a typical ResNet-18/CIFAR-10 model, with the ground truth 
attack target class of backdoor poisoning in red.

\subsection{Layer 5 and 13 Bar Charts of CEPA detection metrics $\|\mu\|$ and $\frac{\sigma}{\|\mu\|}$ on ResNet-18 for CIFAR-10} \label{sec:barcharts_for_layer5_13}
\begin{figure*}[t]
     \centering
     \begin{subfigure}[b]{0.18\textwidth}
         \centering
         \includegraphics[width=\textwidth]{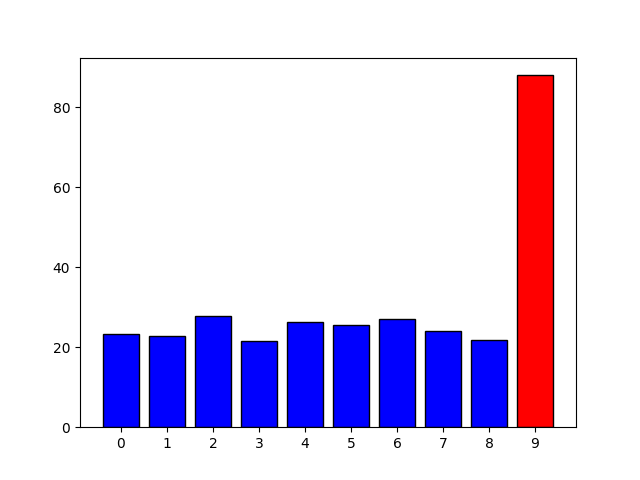}
         \caption{BadNet}
     \end{subfigure}
     \begin{subfigure}[b]{0.18\textwidth}
         \centering
         \includegraphics[width=\textwidth]{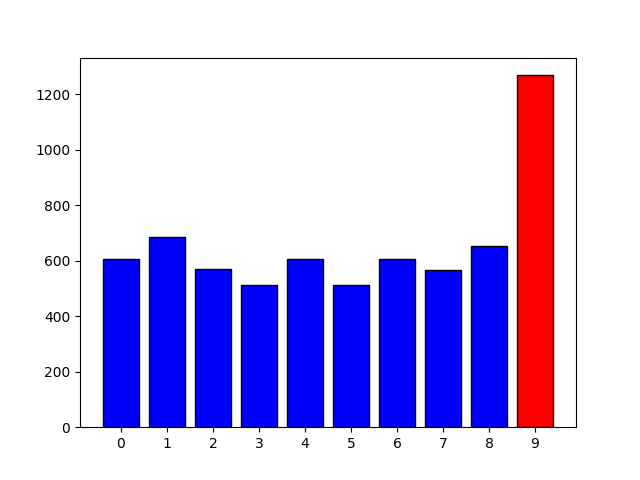}
         \caption{WaNet}
     \end{subfigure}
     \begin{subfigure}[b]{0.18\textwidth}
         \centering
         \includegraphics[width=\textwidth]{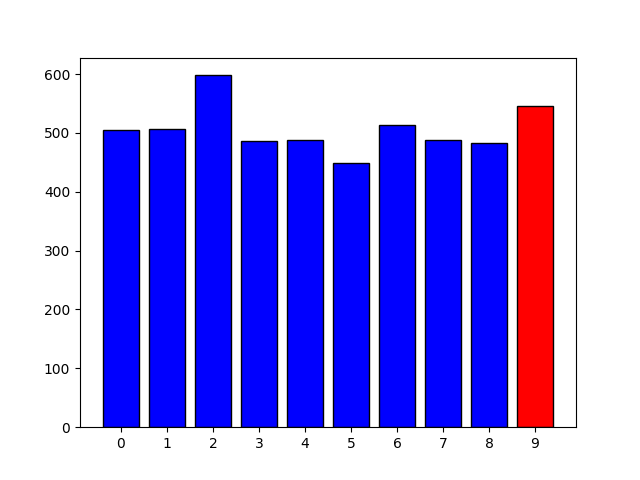}
         \caption{chessboard}
     \end{subfigure}
     \begin{subfigure}[b]{0.18\textwidth}
         \centering
         \includegraphics[width=\textwidth]{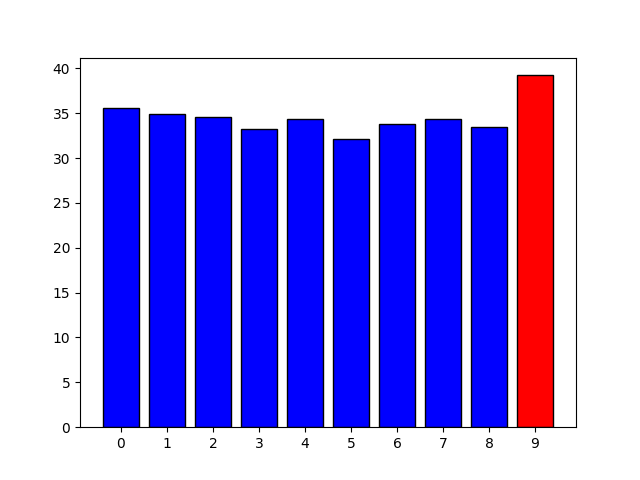}
         \caption{blend}
     \end{subfigure}
     \begin{subfigure}[b]{0.18\textwidth}
         \centering
         \includegraphics[width=\textwidth]{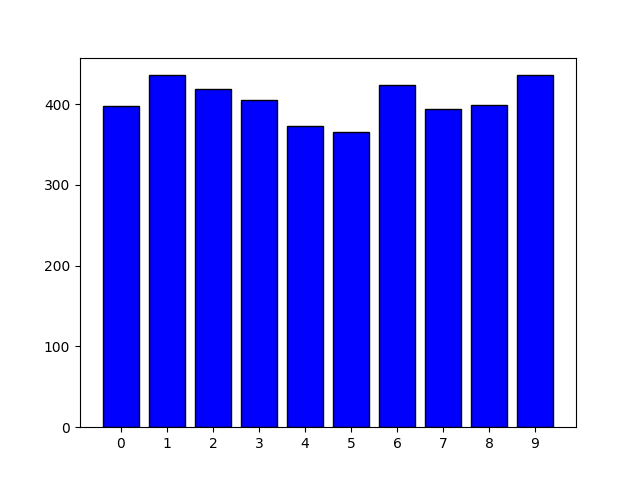}
         \caption{clean}
     \end{subfigure}
        \caption{Layer 5, $\|\mu\|$ values by class, for typical poisoned and clean
        models with true target class in red.}
        \label{fig:layer5-norm}
\end{figure*}

\begin{figure*}[t]
     \centering
     \begin{subfigure}[b]{0.18\textwidth}
         \centering
         \includegraphics[width=\textwidth]{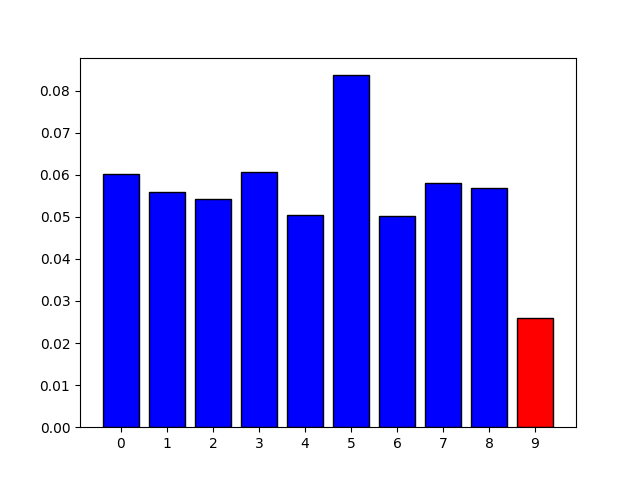}
         \caption{BadNet}
     \end{subfigure}
     \begin{subfigure}[b]{0.18\textwidth}
         \centering
         \includegraphics[width=\textwidth]{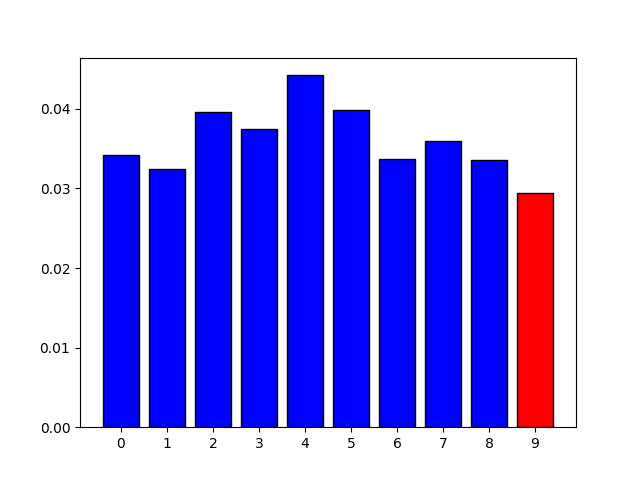}
         \caption{WaNet}
     \end{subfigure}
     \begin{subfigure}[b]{0.18\textwidth}
         \centering
         \includegraphics[width=\textwidth]{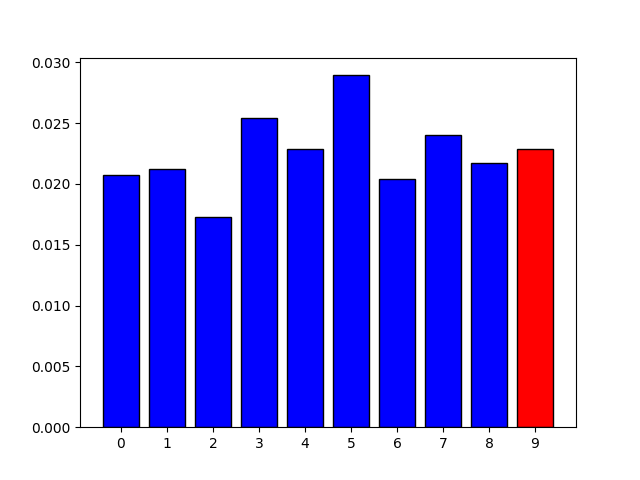}
         \caption{chessboard}
     \end{subfigure}
     \begin{subfigure}[b]{0.18\textwidth}
         \centering
         \includegraphics[width=\textwidth]{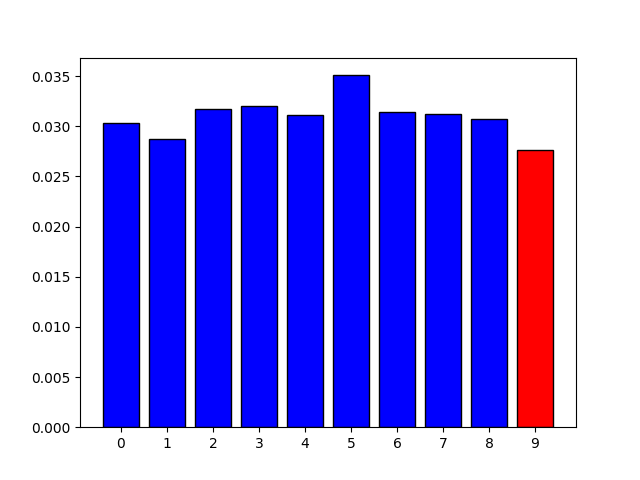}
         \caption{blend}
     \end{subfigure}
     \begin{subfigure}[b]{0.18\textwidth}
         \centering
         \includegraphics[width=\textwidth]{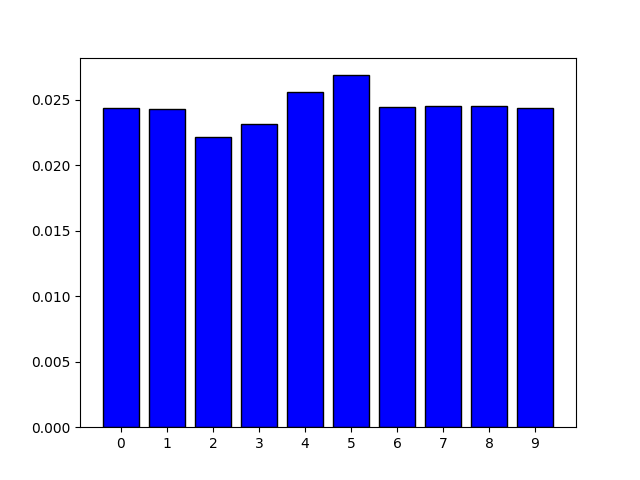}
         \caption{clean}
     \end{subfigure}
        \caption{Layer 5, $\frac{\sigma}{\|\mu\|}$ values by class, for typical poisoned and clean
        models with true target class in red.}
        \label{fig:layer5-consensus}
\end{figure*}

\begin{figure*}[t]
     \centering
     \begin{subfigure}[b]{0.18\textwidth}
         \centering
         \includegraphics[width=\textwidth]{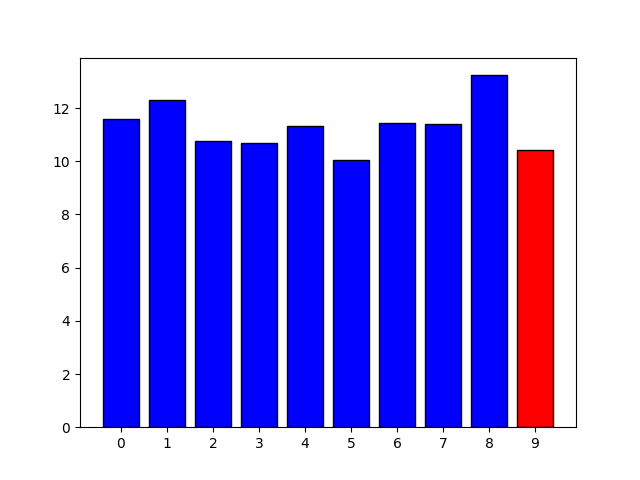}
         \caption{BadNet}
     \end{subfigure}
     \begin{subfigure}[b]{0.18\textwidth}
         \centering
         \includegraphics[width=\textwidth]{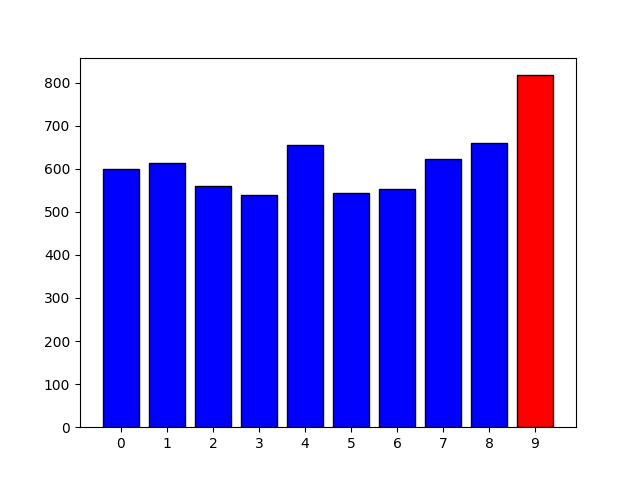}
         \caption{WaNet}
     \end{subfigure}
     \begin{subfigure}[b]{0.18\textwidth}
         \centering
         \includegraphics[width=\textwidth]{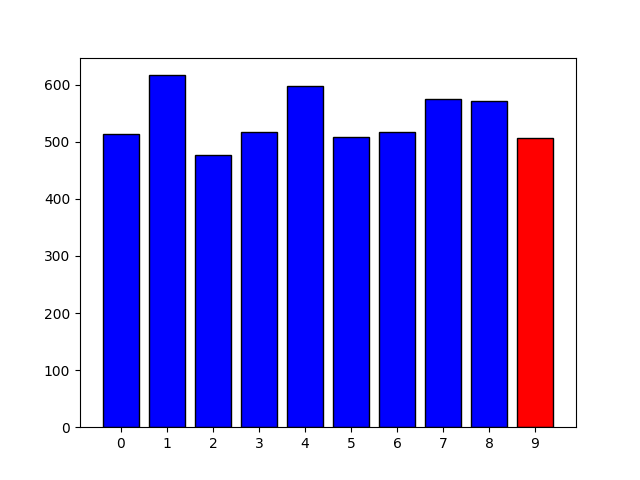}
         \caption{chessboard}
     \end{subfigure}
     \begin{subfigure}[b]{0.18\textwidth}
         \centering
         \includegraphics[width=\textwidth]{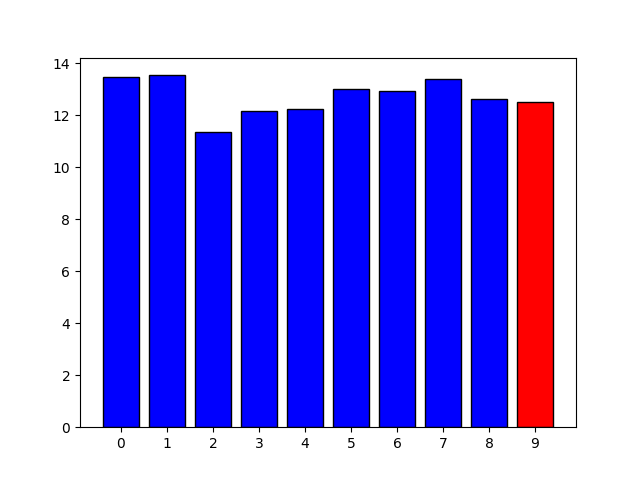}
         \caption{blend}
     \end{subfigure}
     \begin{subfigure}[b]{0.18\textwidth}
         \centering
         \includegraphics[width=\textwidth]{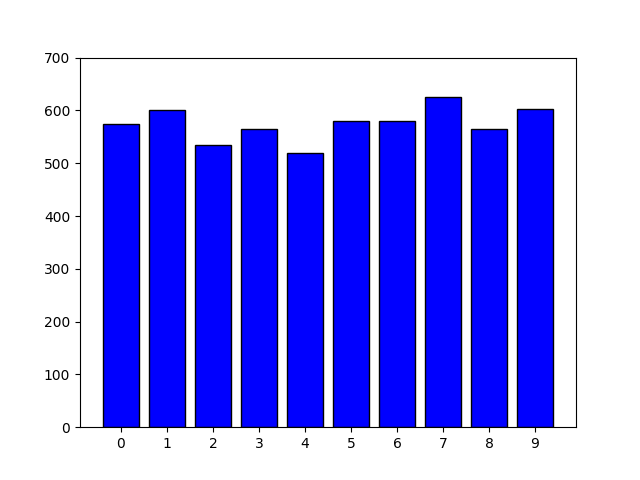}
         \caption{clean}
     \end{subfigure}
        \caption{Layer 13, $\|\mu\|$ values by class, for typical poisoned and clean
        models with true target class in red.}
        \label{fig:layer13-norm}
\end{figure*}

\begin{figure*}[t]
     \centering
     \begin{subfigure}[b]{0.18\textwidth}
         \centering
         \includegraphics[width=\textwidth]{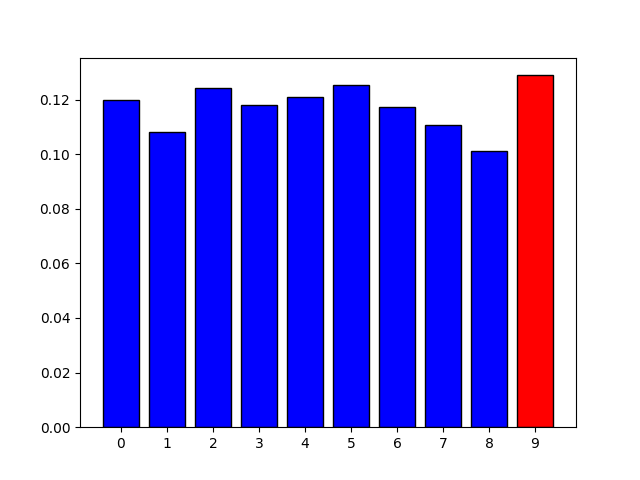}
         \caption{BadNet}
     \end{subfigure}
     \begin{subfigure}[b]{0.18\textwidth}
         \centering
         \includegraphics[width=\textwidth]{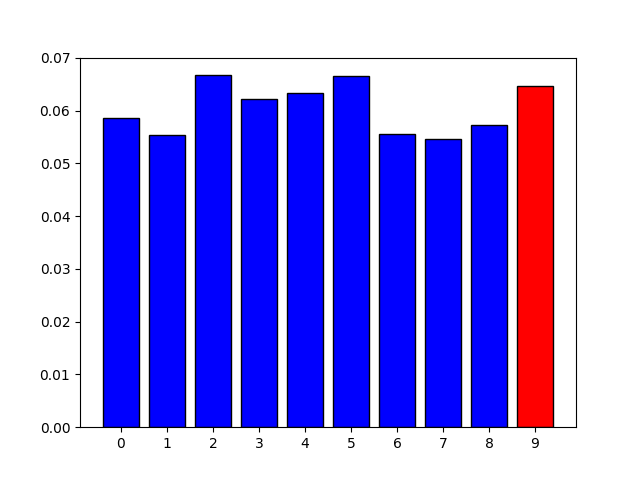}
         \caption{WaNet}
     \end{subfigure}
     \begin{subfigure}[b]{0.18\textwidth}
         \centering
         \includegraphics[width=\textwidth]{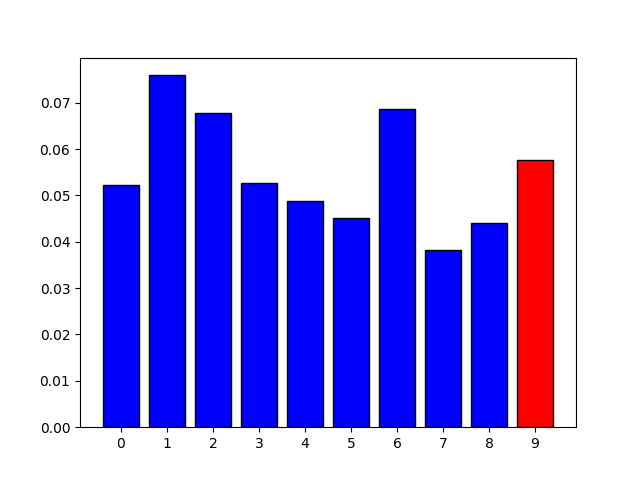}
         \caption{chessboard}
     \end{subfigure}
     \begin{subfigure}[b]{0.18\textwidth}
         \centering
         \includegraphics[width=\textwidth]{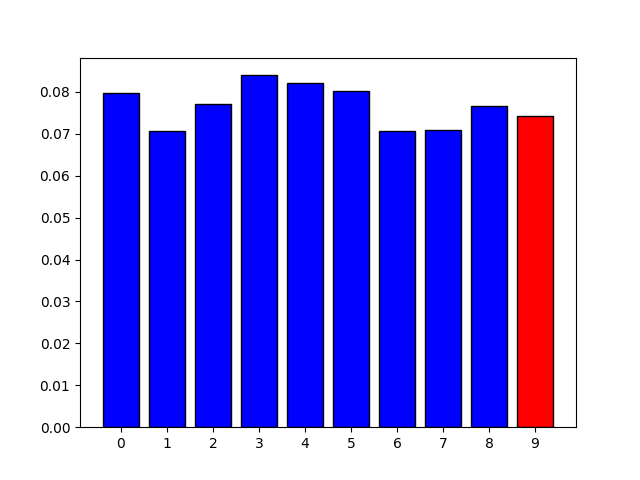}
         \caption{blend}
     \end{subfigure}
     \begin{subfigure}[b]{0.18\textwidth}
         \centering
         \includegraphics[width=\textwidth]{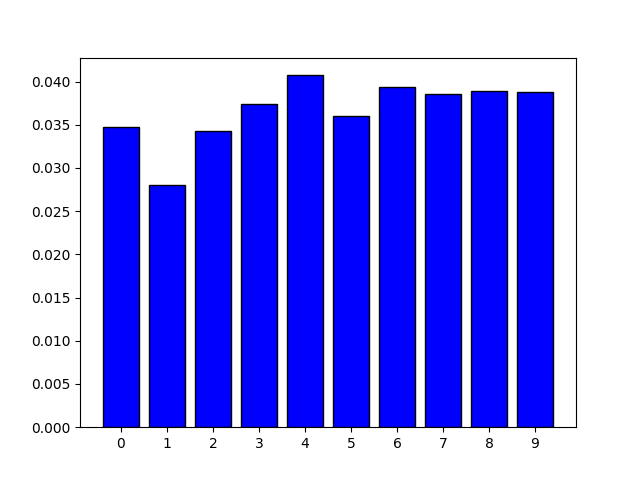}
         \caption{clean}
     \end{subfigure}
        \caption{Layer 13, $\frac{\sigma}{\|\mu\|}$ values by class, for typical poisoned and clean
        models with true target class in red.}
        \label{fig:layer13-consensus}
\end{figure*}

Figures \ref{fig:layer5-norm}-\ref{fig:layer13-consensus} give the bar charts of $\|\mu\|$ and $\frac{\sigma}{\|\mu\|}$ for CEPA applied to layers 5 and 13 of the ResNet-18/CIFAR-10 model. Collectively, the bar charts demonstrate the superiority of layer 9 in detection, as it successfully identifies all attacks without producing any false alarms on the clean model. 




\subsection{Detection based on cosine similarity of CEPA optimized perturbation $\delta$s}\label{sec:cosine-similarity}

\begin{figure*}[t]
    \centering
     \begin{subfigure}[b]{0.18\textwidth}
         \centering
         \includegraphics[width=\textwidth]{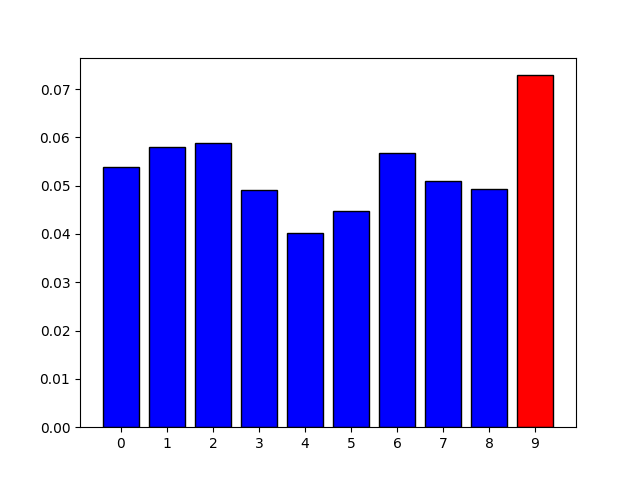}
         \caption{BadNet}
         \label{fig:badnet}
     \end{subfigure}
     \begin{subfigure}[b]{0.18\textwidth}
         \centering
         \includegraphics[width=\textwidth]{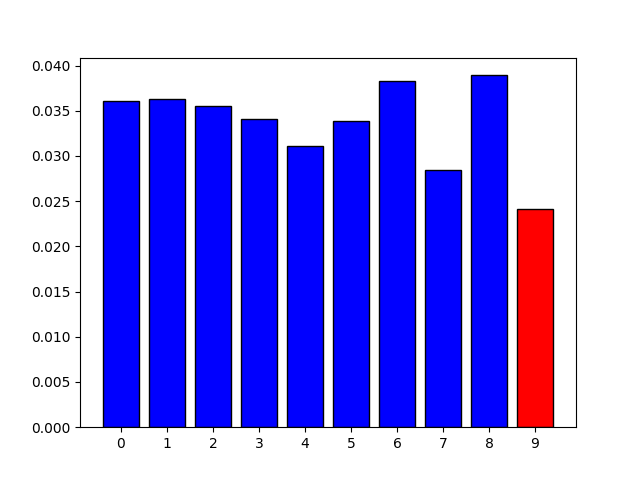}
         \caption{WaNet}
         \label{fig:WaNet}
     \end{subfigure}
     \begin{subfigure}[b]{0.18\textwidth}
         \centering
         \includegraphics[width=\textwidth]{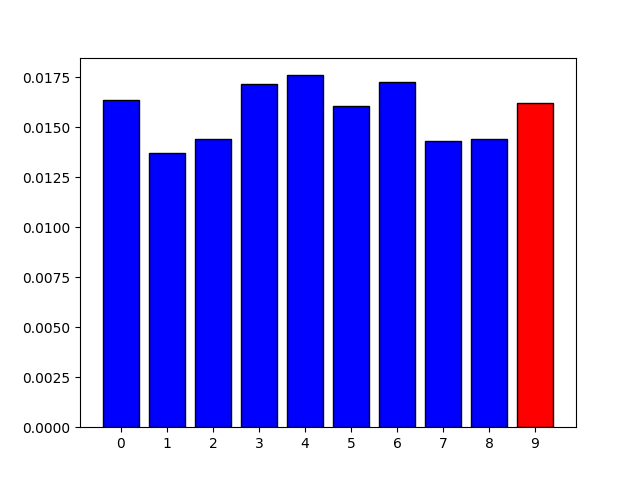}
         \caption{chessboard}
         \label{fig:chessboard}
     \end{subfigure}
    \begin{subfigure}[b]{0.18\textwidth}
         \centering
         \includegraphics[width=\textwidth]{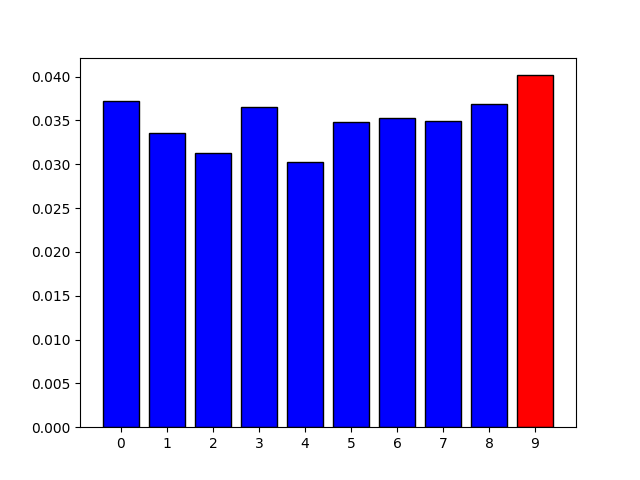}
         \caption{blend}
         \label{fig:blend}
     \end{subfigure}
     \begin{subfigure}[b]{0.18\textwidth}
         \centering
         \includegraphics[width=\textwidth]{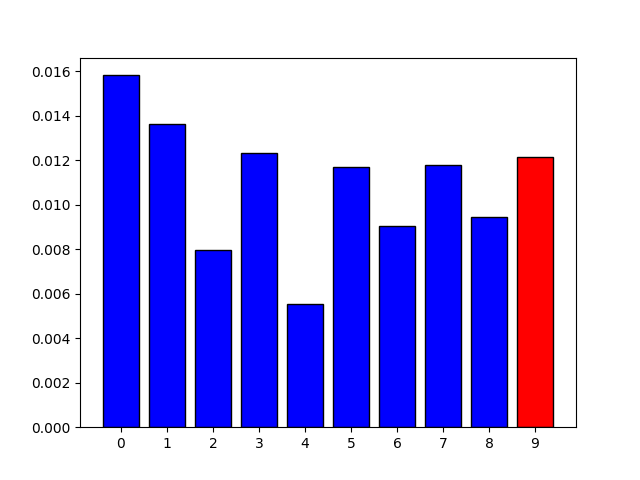}
         \caption{clean}
         \label{fig:clean}
     \end{subfigure}
    \caption{Layer 9, mean pairwise cosine similarity of the optimized perturbation $\delta$s by class, for typical poisoned and clean models with true target class in red.}
    \label{fig:delta-cos-sim}
\end{figure*}

Some prior work on backdoor detection has suggested the use of cosine similarity 
as a detection statistic, e.g., \cite{RWang20}.
In Figure \ref{fig:delta-cos-sim}, for a typical ResNet-18/CIFAR-10 model we plot the mean (pairwise) cosine similarity of the optimized $\delta$s for each putative target class. From the bar charts, one can see that there is no common trend (either high or low similarity) observed for the true target class (9) across attacks.  
Therefore, mean cosine similarity is not effective as a detection statistic.
This is in contrast to$\|\Bar{\delta}\|$, $\|\mu\|$ and  $\frac{\sigma}{\|\mu\|}$ (inspired by the backdoor overfitting phenomenon recognized in \cite{backdoor-universal}), which, when used collectively are effective detection statistics.

\subsection{Detection performance for different misclassification thresholds}

As discussed before, CEPA only truly possesses one hyperparameter: the misclassification threshold $\pi_0$. However, our experiments show that CEPA is largely insensitive to this hyperparameter. In tables \ref{tab:MAD-pi-8} and \ref{tab:MAD-pi-5}, we changed $\pi_0$ to 0.8 and 0.5, respectively. CEPA applied on layer 9 still demonstrates good detection performance across attacks, with no false positives for the clean model. The only exception is for the blend attack when $\pi_0=0.5$, where all three metrics failed to detect.
\begin{table}
    \centering
    \begin{tabular}{c|ccccc}
    \toprule
     layer, crit. & BN & WN & CB & Bl. & Clean  \\ 
     \hline
     9,  $\|\textstyle\Bar{\delta}\|$   
     & 0.92(8) & 1.55(3) & 1.61(5) & 1.75(8) & 0.78(8) \\ 
     9,  $\|\textstyle\mu\|$   
     & 16.23(9) & 4.74(9) & 3.79(9) & 3.10(9) & 0.89(6) \\ 
     9,  $\frac{\sigma}{\|\mu\|}$   
     & 7.07(9) & 0.84(9) & 1.18(9) & 1.54(1) & 1.33(1) \\
    \bottomrule
    \end{tabular}
    \caption{CEPA MAD scores with $\pi_0=0.8$. The score represents the largest MAD value among all potential target classes. Along with each score, the class achieving that score is indicated in the parenthesis. The attacks are
    BadNet (BN), WaNet (WN), chessboard (CB), and Blend (Bl.).}
    \label{tab:MAD-pi-8}
\end{table}
\begin{table}
    \centering
    \begin{tabular}{c|ccccc}
    \toprule
     layer, crit. & BN & WN & CB & Bl. & Clean  \\ 
     \hline
     9,  $\|\textstyle\Bar{\delta}\|$   
     & 1.08(8) & 0.77(3) & 1.18(3) & 1.63(8) & 1.14(8) \\ 
     9,  $\|\textstyle\mu\|$   
     & 6.01(9) & 3.09(9) & 3.09(9) & 3.11(6) & 0.94(1) \\ 
     9,  $\frac{\sigma}{\|\mu\|}$   
     & 2.87(9) & 0.81(1) & 0.99(9) & 0.85(6) & 1.77(1) \\
    \bottomrule
    \end{tabular}
    \caption{CEPA MAD scores with $\pi_0=0.5$. }
    \label{tab:MAD-pi-5}
\end{table}

\subsection{Inversion visualization for different $\lambda_1$ and $\lambda_2$}
\label{sec:inv-different-lambda}

\begin{figure}[!ht]
    \centering
    \begin{subfigure}[b]{0.28\columnwidth}
         \includegraphics[width=\textwidth]{fig/BadNet_GT.png}
         \caption{}
         \label{fig:badnet-gt1}
     \end{subfigure}
    \begin{subfigure}[b]{0.28\columnwidth}
        \includegraphics[width=\textwidth]{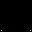}
         \caption{}
         \label{fig:BadNet_inv_0001_5}
     \end{subfigure}
     \begin{subfigure}[b]{0.28\columnwidth}
        \includegraphics[width=\textwidth]{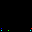}
         \caption{}
         \label{fig:BadNet_inv_00001_25}
     \end{subfigure}
         \caption{For the BadNet attack: (a) the ground truth perturbation, (b) the average estimated perturbation $|\mathcal{D}|^{-1}\sum_{x\in \mathcal{D}}\hat{\delta}_x$ with $\lambda_1=0.0001, \lambda_2=0.5$, and (c) the average estimated perturbation with $\lambda_1=0.00001, \lambda_2=2.5$.}
         \label{fig:visual_BadNet_different_lambda}
         \vspace{-0.1in}
         \end{figure}

The inversion patterns obtained from CEPA are not sensitive to the particular choices made for $\lambda_1$ and $\lambda_2$. In fact, we have found that any fixed positive
$\lambda_1$ and $\lambda_2$ that result in small optimized perturbations $\delta$ (e.g.,
average norm smaller than 1) yield good estimates of the
backdoor pattern.

To support this,
in Figure \ref{fig:visual_BadNet_different_lambda}(b), we display the CEPA inversion result for the example of a BadNet attacked ResNet-18/CIFAR-10 model, where we changed $\lambda_1$ and $\lambda_2$ to $0.0001$ and $0.5$, respectively. In Figure \ref{fig:visual_BadNet_different_lambda}(c), we changed $\lambda_1$ and $\lambda_2$ to $0.00001$ and $2.5$, respectively. Figure \ref{fig:visual_BadNet_different_lambda}(a) shows the ground truth backdoor trigger. One can observe from these figures that varying the $\lambda$ values does not significantly affect the estimated pattern -- it remains indicative of the ground truth trigger pattern and its location.
\subsection{BadNet against VGG-16 for CIFAR-100}\label{sec:CIFAR-100}
 

\begin{table}[ht]
\centering
\begin{tabular}{c|ccc}
\toprule
{layer$\backslash$ criterion} & $\|\Bar{\delta}\|$ & $\|\mu\|$ & $\frac{\sigma}{\|\mu\|}$ \\ \hline
  3  & 2.04(78) & 33.72(0) & 7.44(0) \\ \hline
  8 & 1.73(78)  & 22.40(0) & 6.54(0) \\ 
  \bottomrule
\end{tabular}
\caption{CEPA MAD scores for BadNet on VGG-16/CIFAR-100.}
\label{tab:MAD-100}
\end{table}

\begin{table}[ht]
\centering
\begin{tabular}{c|ccc}
\toprule
{layer$\backslash$ criterion} & $\|\Bar{\delta}\|$ & $\|\mu\|$ & $\frac{\sigma}{\|\mu\|}$ \\ \hline
  3  & 1.70(22) & 1.90(58) & 1.61(41) \\ \hline
  8 & 1.88(78) & 2.53(71) & 1.78(71) \\ 
  \bottomrule
\end{tabular}
\caption{CEPA MAD scores for the clean model on VGG-16/CIFAR-100.}
\label{tab:MAD-100-clean}
\end{table}

In Table \ref{tab:MAD-100},
we show the CEPA detection statistics' MAD scores for a poisoned model at layers 3 and 8. Both layers (easily) detect the backdoor (with target class 0). Note also that there is a (barely) false positive detection on class 78 using $\|\Bar{\delta}\|$ at layer 3. There were no false positives in both layers for the clean/unpoisoned CIFAR-100 model, as all metrics' MAD scores did not exceed the given thresholds, shown in Table \ref{tab:MAD-100-clean}.

\subsection{Adaptive additive attack on ResNet-18 for CIFAR-10}
\label{sec:adaptive}
Here, we consider an additive backdoor incorporation mechanism,
and a scenario wherein the attacker optimizes the backdoor pattern given knowledge
of our detection defense.  In particular,
assume the attacker is the training authority and
trains the model and, jointly, optimizes over the backdoor pattern,
to {\it minimize} consensus (to maximize the variance of the
embedded perturbation corresponding to the backdoor).
That is, for the training set 
$\mathcal{X}$ and backdoor-poisoned
training subset $\mathcal{X}_B \subset \mathcal{X}$,
the
training objective is
the cross-entropy loss on
$\mathcal{X}$
 minus parameter
$\gamma>0$ times the variance about the embedded perturbations induced by
the ground-truth backdoor-triggered patterns in the poisoned training subset,
$(x+\delta) \in\mathcal{X}_{\rm B}\subset \mathcal{X}$:
\begin{eqnarray*}
\lefteqn{
-\sum_{(x,y)\in\mathcal{X}}\log \hat{p}(y|x)} && \\
& & -~\frac{\gamma}{|\mathcal{X}_{\rm B}|} \sum_{(x+\delta) \in \mathcal{X}_{\rm B}}
\| f(x+\delta)-f(x) - \mu_{\rm B} \|^2, 
\end{eqnarray*}
where $\delta$ is the (common) additive backdoor pattern and
where
$$\mu_{\rm B} =
\frac{1}{|\mathcal{X}_{\rm B}|} \sum_{(x+\delta) \in \mathcal{X}_{\rm B}}
(f(x+\delta)-f(x)).$$
This objective function is jointly minimized with respect to both the DNN's model
parameters and the additive backdoor pattern, $\delta$.

Based on this approach, we launched an adaptive attack on a Resnet-18 model 
trained on the CIFAR-10 dataset, with 
100 training epochs and 
a learning rate of 0.001. 
In one experiment, the target was class 9, the adaptive attack parameter $\gamma$ was set to 1,
the embedding layer considered by the 
attacker was layer 5 of the Resnet-18 model, and the poisoning rate
was high at 0.5 (50\%) --
under the adaptive attack, a high poisoning rate was needed to achieve a reasonable
attack success rate (ASR), which in this case was 0.75.
The clean-data accuracy (ACC) of the poisoned model was 0.83.
Subsequently, we implemented CEPA detection on the $9^{\rm th}$ layer of the poisoned Resnet-18 model. 
The target class, class 9, was successfully detected: the largest MAD score,  
amongst all candidate target
classes was 1.64 from class 8, 8.73 from class 9, and 4.75 from class 9, respectively, for the 
$\|\Bar{\delta}\|$, $\|\mu\|$, and $\frac{\sigma}{\|\mu\|}$  detection statistics.

\subsection{Ablation Studies}
\label{sec:ablation}
\begin{table}
    \centering
    \begin{tabular}{c|ccccc}
    \toprule
     layer, crit. & BN & WN & CB & Bl. & Clean  \\ 
     \hline
     9,  $\|\textstyle\Bar{\delta}\|$   
     & 1.86(9) & 7.37(3) & 3.45(7) & 0.98(8) & 2.89(3) \\ 
     9,  $\|\textstyle\mu\|$   
     & 1.97(6) & 3.67(9) & 1.94(9) & 1.33(7) & 0.85(1) \\ 
     9,  $\frac{\sigma}{\|\mu\|}$   
     & 1.35(6) & 1.12(0) & 1.29(6) & 3.39(7) & 1.20(6) \\
    \bottomrule
    \end{tabular}
    \caption{MAD scores obtained after optimizing objective \ref{objective} without the embedded consensus constraint.}
    \label{tab:MAD-ablation}
\end{table}

For detection, the consensus constraint on the embedded features is crucial for detecting backdoors; 
recall objective \ref{objective}. In Table \ref{tab:MAD-ablation}, 
we present the MAD scores obtained after optimizing \ref{objective} {\it without} 
the constraint on embedded consensus. That is, we are optimizing $\delta_x$ only seeking to increase the misclassification rate to the target class. 
The MAD scores $\|\Bar{\delta}\|$, $\|\mu\|$, and $\frac{\sigma}{\|\mu\|}$ are 
not indicative of the true target class, for all attacks -- the one exception is for $\|\mu\|$ on WaNet. 

\begin{figure*}[ht]
\centering
    \begin{subfigure}[b]{0.28\columnwidth}
         \includegraphics[width=\textwidth]{fig/BadNet_GT.png}
         \caption{}
     \end{subfigure}
    \begin{subfigure}[b]{0.28\columnwidth}
        \includegraphics[width=\textwidth]{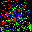}
         \caption{}
     \end{subfigure}
     \caption{Inverted trigger optimized 
     without a constraint on the input perturbation norm for the BadNet attack: 
     (a) the ground truth perturbation, and (b) the average estimated perturbation $|\mathcal{D}|^{-1}\sum_{x\in \mathcal{D}}\hat{\delta}_x$.}
     \label{fig:badnet_ablation}
\end{figure*}

For reverse engineering, the constraint on input perturbation norm 
is needed for proper inverted estimation of perturbations. 
If we set $\lambda_2=0$ (i.e. no constraint on the perturbation norm), 
the resulting $\|\delta\|$ will be too large to follow the backdoor "shortcut"; 
in this case the inverted perturbation tends to be $\sim$ more consistent with a TTE attack and, thus,
is an inaccurate estimate of the backdoor pattern. 
See Figure \ref{fig:badnet_ablation} for an example of a BadNet attack -- an accurate inverted pattern for Badnet should be spatially localized,
but in this case a global pattern is estimated.

\end{document}